\documentclass[letterpaper,titlepage,10pt]{article}

\usepackage{hyperref}

\usepackage{amssymb,amsmath,amsfonts}
\usepackage{epsfig}

\def\clock{{\count0=\time
           \divide\count0 60
           \ifnum\count0<10 0\fi\the\count0
           \multiply\count0 -60 \advance\count0 \time
           :\ifnum\count0<10 0\fi \the\count0
         }}
\newcommand{\timestamp}{{\small\vbox{\hbox{\tt\jobname.tex}
\hbox{\the\day/\the\month/\the\year, \clock}}}}

\newcommand{\CL}{\mathcal{L}}
\newcommand{\CO}{\mathcal{O}}
\newcommand{\CN}{\mathcal{N}}

\newcommand{\CH}{\mathcal{H}}

\newcommand{\Z}{\mathbb{Z}}
\newcommand{\C}{\mathbb{C}}
\newcommand{\R}{\mathbb{R}}

\newcommand{\nn}{\nonumber}

\newcommand{\spa}{\ , \ \ }

\newcommand{\ds}{\displaystyle}

\newcommand{\tr}{\mathop{{\rm Tr}}}

\newcommand{\ads}{\mbox{AdS}}

\begin{document}

\begin{titlepage}

\rightline{\vbox{\small\hbox{\tt UUITP-14/08} }}
 \vskip 2 cm

\centerline{\LARGE \bf Finite-size corrections in the $SU(2)\times
SU(2)$ sector of} \vskip 0.2cm \centerline{\LARGE \bf type IIA
string theory on $\ads_4\times \C P^3$} \vskip 1.7cm

\centerline{\large {\bf Davide Astolfi$\,^{1}$}, {\bf Valentina
Giangreco M. Puletti$\,^{2}$}, {\bf Gianluca Grignani$\,^{1}$},}
\vskip 0.2cm
\centerline{\large {\bf Troels Harmark$\,^{3}$} and
{\bf Marta Orselli$\,^{3}$} }

\vskip 0.5cm

\begin{center}
\sl $^1$ Dipartimento di Fisica, Universit\`a di Perugia,\\
I.N.F.N. Sezione di Perugia,\\
Via Pascoli, I-06123 Perugia, Italy\\
\vskip 0.4cm
\sl $^2$ Department of Physics and Astronomy, Uppsala University,\\
SE-75108 Uppsala, Sweden \vskip 0.4cm
\sl $^3$ The Niels Bohr Institute  \\
\sl  Blegdamsvej 17, 2100 Copenhagen \O , Denmark \\
\end{center}
\vskip 0.5cm

\centerline{\small\tt astolfi@pg.infn.it,
valentina.giangreco@fysast.uu.se, grignani@pg.infn.it,}
\centerline{\small\tt harmark@nbi.dk, orselli@nbi.dk}

\vskip 1.5cm

\centerline{\bf Abstract} \vskip 0.2cm \noindent We consider
finite-size corrections in the $SU(2)\times SU(2)$ sector of type
IIA string theory on $\ads_4\times \C P^3$, which is the string dual
of the recently constructed $\CN=6$ superconformal Chern-Simons
theory of Aharony, Bergman, Jafferis and Maldacena (ABJM theory).
The string states we consider are in the $\R \times S^2\times S^2$
subspace of $\ads_4\times \C P^3$ with an angular momentum $J$ on
$\C P^3$ being large. We compute the finite-size corrections using
two different methods, one is to consider curvature corrections to
the Penrose limit giving an expansion in $1/J$, the other by
considering a low energy expansion in $\lambda' = \lambda /
J^2$ of the string theory sigma-model, $\lambda$ being the 't Hooft
coupling of the dual ABJM theory. For both methods there are
interesting issues to deal with. In the near-pp-wave method there is
a $1/\sqrt{J}$ interaction term for which we use zeta-function
regularization in order to compute the $1/J$ correction to the
energy. For the low energy sigma-model expansion we have to take
into account a non-trivial coupling to a non-dynamical transverse
direction. We find agreement between the two methods. At order
$\lambda'$ and $\lambda'^2$, for small $\lambda'$, our results are
analogous to the ones for the $SU(2)$ sector in type IIB string
theory on $\ads_5\times S^5$. Instead at order $\lambda'^3$ there
are interactions between the two two-spheres. We compare our results
with the recently proposed all-loop Bethe ansatz of Gromov and
Vieira and find agreement.

%\vskip 0.5cm \leftline{\timestamp}

\end{titlepage}

\pagestyle{plain} \setcounter{page}{1}

\tableofcontents

%%%%%%%%%%%%%%%%%%%%%%%%%%%%%%%%%%%%%%%%%%%%%%%%%%%%%%%%%%%%%
\section{Introduction and summary}
\label{sec:intro}

Recently Aharony, Bergman, Jafferis and Maldacena proposed a new
exact duality between gauge theory and string theory
\cite{Aharony:2008ug} based on earlier work on superconformal
Chern-Simons theories
\cite{Schwarz:2004yj}.%
\footnote{For papers considering the ABJM theory see
\cite{Benna:2008zy}.} The new duality is between a three-dimensional
$\CN=6$ superconformal Chern-Simons theory (ABJM theory), and type
IIA string theory on $\ads_4\times \C P^3$. ABJM theory has
$SU(N)\times SU(N)$ gauge symmetry with Chern-Simons like kinetic
terms at level $k$ and it is weakly coupled when the 't Hooft
coupling $\lambda = N/k$ is small. Instead type IIA string theory on
$\ads_4\times \C P^3$ is a good description when $1 \ll \lambda \ll
k^4$.

Subsequently it was found in \cite{Minahan:2008hf,Gaiotto:2008cg}
that the $SU(4)$ R-symmetry sector of ABJM theory is integrable at two-loop
order. In particular one can
consider the $SU(2)\times SU(2)$ sector of $SU(4)$. The operators in
this sector are constructed from the single-trace operators of the
form
\begin{equation}
\label{su2op}
\tr ( A_{i_1} B_{j_1} A_{i_2} B_{j_2} \cdots A_{i_J} B_{j_J} )
\end{equation}
with $A_{1,2}$ and $B_{1,2}$ transforming in the $(1/2,0)$ and
$(0,1/2)$ of $SU(2)\times SU(2)$, respectively, and all scalars
being in the bifundamental representation of $SU(N)\times SU(N)$. It
was found in \cite{Minahan:2008hf,Gaiotto:2008cg} that this sector is
described by two separate Heisenberg $XXX_{1/2}$ spin chains, with
$A_{1,2}$ corresponding to the up and down spins in the first
Heisenberg chain, and $B_{1,2}$ to the second Heisenberg chain, the
only interaction between them being the zero total momentum
constraint of the magnons.

In \cite{Grignani:2008is} the $SU(2)\times SU(2)$ sector was studied
from the string theory side. The $SU(2)\times SU(2)$ sector
corresponds on the string theory side to considering an $\R\times
S^2 \times S^2$ subspace of the $\ads_4\times \C P^3$ background.
The $SU(2)\times SU(2)$ sector was approached by taking a low-energy
sigma-model limit, with the result that at leading order the
sigma-model action is that of two Landau-Lifshitz models added
together. This is consistent with what one finds on the gauge theory
side. Furthermore, a Penrose limit approaching the $SU(2)\times
SU(2)$ sector was considered in \cite{Grignani:2008is} (see also
\cite{Nishioka:2008gz,Gaiotto:2008cg}) and a new Giant magnon
solution was found in the $SU(2)\times SU(2)$ sector
\cite{Grignani:2008is,Grignani:2008te} \footnote{For the giant magnon solution in $AdS_5\times S^5$ see for example~\cite{Hofman:2006xt,Arutyunov:2006gs,Astolfi:2007uz}.} (see also
\cite{Gaiotto:2008cg}). Combining these studies it was found that a
magnon in the $SU(2)\times SU(2)$ sector has a dispersion relation
that depends non-trivially on the coupling
\cite{Gaiotto:2008cg,Grignani:2008is}
\begin{equation}
\label{dispgen} \Delta = \sqrt{ \frac{1}{4} + h(\lambda) \sin^2
\Big( \frac{p}{2} \Big) } \spa h(\lambda) = \left\{ \begin{array}{c}
\ds 4\lambda^2 + \CO ( \lambda^4 )
 \ \mbox{for} \ \lambda \ll 1 \\[4mm] \ds 2 \lambda + \CO ( \sqrt{\lambda} )
 \ \mbox{for} \ \lambda \gg 1 \end{array} \right.
\end{equation}
where the weak coupling result is from \cite{Minahan:2008hf,Gaiotto:2008cg}.

Very recently a proposal for an all-loop Bethe ansatz for the
$\ads_4 / \mbox{CFT}_3$ duality was put forward
in~\cite{Gromov:2008qe}. This proposal combines the full
$OSp(2,2|6)$ superconformal symmetry with the results on
integrability of ABJM theory found at weak coupling
\cite{Minahan:2008hf,Gaiotto:2008cg}, the interpolating dispersion
relation \eqref{dispgen} of \cite{Gaiotto:2008cg,Grignani:2008is}
and the study of integrability on the string theory side
\cite{Arutyunov:2008if,Stefanski:2008ik,Fre:2008qc,Gromov:2008bz}.
The proposal utilizes many ingredients of the all-loop proposal for
$\CN=4$ super-Yang-Mills theory (SYM)
\cite{Beisert:2003ys,Arutyunov:2004vx,Beisert:2006ez}.

In this paper we continue the study of integrability of the $\ads_4 / \mbox{CFT}_3$
duality by computing the finite-size corrections to string states in the $SU(2)\times SU(2)$
sector of type IIA string theory on $\ads_4\times \C P^3$ with a large angular momentum $J$ on $\C P^3$.
The string states are dual to single-trace operators of the form \eqref{su2op} in ABJM theory with $2J$
being the number of complex scalars in the operator.%
\footnote{We pick the three Cartan generators of the $SU(4)$ R-symmetry (or the $SU(4)$
symmetry of $\C P^3$) $R_1$, $R_2$ and $R_3$  such that $J = - R_3$, $S_z^{(1)} = (R_1-R_2)/2$
and $S_z^{(2)}=(R_1+R_2)/2$ where $S_z^{(1,2)}$ are the Cartan generators for the two $SU(2)$'s \cite{Grignani:2008is}.}
We compute the finite-size corrections using
two different methods. The first method is to consider curvature corrections to
the Penrose limit of \cite{Grignani:2008is} giving an expansion in $1/J$. The second method is to make a low energy expansion in $\lambda' \equiv \lambda /
J^2$ of the string theory sigma-model, expanding around the $SU(2)\times SU(2)$ sigma-model limit of \cite{Grignani:2008is}.

For the curvature corrections to the Penrose limit we follow the
pioneering approach of  \cite{Callan:2003xr,Callan:2004uv} in which
curvature corrections to the BMN pp-wave \cite{Berenstein:2002jq}
were considered for type IIB string theory on $\ads_5\times S^5$.
For simplicity we focus on string states in the $SU(2)\times SU(2)$
sector. We compute the $1/J$ correction to the energy of two
different string states: $|s\rangle$ which is a two-oscillator state
in the first $SU(2)$ and $|t\rangle$ which is a two-oscillator state
with one oscillator in each of the $SU(2)$'s. The computation
involves a new feature compared with that of
\cite{Callan:2003xr,Callan:2004uv}, namely that a $1/\sqrt{J}$
curvature correction appears in the Hamiltonian involving a
transverse direction. This $1/\sqrt{J}$ correction appears as a
second order correction at order $1/J$ giving a finite contribution
to the energy after using zeta-function regularization.

For the state $|s\rangle$ we find the following energy
\begin{equation}\label{Eaa}
E_s=2\sqrt{ \frac{1}{4} + 2\pi^2 n^2\lambda'  }
-1+\frac{\lambda'}{J}\frac{4 \pi^2 n^2}{\frac{1}{4}+2 \pi^2
n^2\lambda'}\left(\sqrt{ \frac{1}{4} + 2\pi^2n^2\lambda'  }-2\pi^2
n^2\lambda'\right)
\end{equation}
where $n$ is the oscillator number. For the state $|t\rangle$ we
find
\begin{equation}\label{eba}
E_t=2\sqrt{ \frac{1}{4} + 2\pi^2 n^2\lambda'  }
-1+\frac{\lambda'}{J}\frac{4 \pi^2 n^2}{\frac{1}{4}+2 \pi^2
n^2\lambda'}\left(\sqrt{ \frac{1}{4} + 2\pi^2n^2\lambda'  }-2\pi^2
n^2\lambda'-\frac{1}{2}\right)
\end{equation}
Here $E = \Delta - J$. The computation of these energies is one of
the main results of this paper.

Expanding the energies \eqref{Eaa} and \eqref{eba} of the two states
$|s\rangle$ and $|t\rangle$ we find that at order $\lambda'$ and
$\lambda'^2$ the $1/J$ correction is what one would expect from
knowing the $1/J$ correction to the $SU(2)$ sector of type IIB
string theory on $\ads_5\times S^5$. For the state $|t\rangle$ this
entails that there is no interaction between the two $SU(2)$'s to
this order which means that there are no $1/J$ corrections at order
$\lambda'$ and $\lambda'^2$.

At order $\lambda'^3$ new interesting effects in the finite-size
corrections appear. Most interestingly, the two $SU(2)$'s start to
interact, and we get a non-zero $1/J$ correction to the string state
$|t\rangle$. In particular, this means that the finite-size
correction starts to deviate at this order from what one could
naively expect from the $SU(2)$ sector in $\ads_5\times S^5$.

Our second method to consider finite-size corrections consists in
making a low-energy expansion of the sigma-model on $\ads_4\times \C
P^3$, with the energy $\Delta-J$ being small. This is an expansion
in $\lambda' = \lambda/J^2$ around the $SU(2)\times SU(2)$
sigma-model limit of \cite{Grignani:2008is}. This method builds on
the analogous low-energy sigma-model limit for the $SU(2)$ sector in
$\ads_5\times S^5$ \cite{Kruczenski:2003gt,Kruczenski:2004kw}. In
parallel to the curvature correction, this computation also involves
a new feature in comparison to
\cite{Kruczenski:2003gt,Kruczenski:2004kw}. The new feature is that
a field corresponding to a transverse direction has a non-trivial
coupling to the fields of the $SU(2)\times SU(2)$ sector even though
the field becomes non-dynamical in the $\lambda' \rightarrow 0$
limit.

To first order in $\lambda'$ we have the result of
\cite{Grignani:2008is} that the sigma-model is two Landau-Lifshitz
models added together without any interaction terms. To second order
in $\lambda'^2$ we find again no interaction terms and the
sigma-model corresponds to two copies of the sigma-model found in
the $SU(2)$ sector of $\ads_5\times S^5$. At third order in
$\lambda'^3$ new interesting effects appear and we get both
interaction terms and new non-trivial terms for each of the
$SU(2)$'s. We check for the two string states that our results are
consistent with the results found from the curvature corrections to
the Penrose limit by comparing with the energies \eqref{Eaa} and
\eqref{eba} expanded up to third order in $\lambda'$.

Finally, we compare our results for the finite-size corrections to
string states in the $SU(2)\times SU(2)$ sector to the newly
proposed all loop Bethe ansatz \cite{Gromov:2008qe}. We write down
the explicit Bethe ansatz for the $SU(2)\times SU(2)$ sector that
results from their proposal. Using this we compute the $1/J$ finite
size corrections to the two string states up to order $\lambda'^8$,
for small $\lambda'$. Amazingly, we find perfect agreement up to
that order. This constitutes a rather non-trivial check of the
proposal of \cite{Gromov:2008qe}.

%%%%%%%%%%%%%%%%%%%%%%%%%%%%%%%%%%%%%%%%%%%%%%%%%%%%%%%%%%%%%
\section{Preliminaries}
\label{sec:prem}

ABJM theory is an $\CN=6$ superconformal Chern-Simons theory with
gauge group $SU(N)\times SU(N)$ and level $k$. For $1 \ll \lambda
\ll k^4$ it is well-described by type IIA string theory on $\ads_4
\times \C P^3$ \cite{Aharony:2008ug}. The $\ads_4\times \C P^3$
background has the metric
\begin{equation}
\label{adscp} ds^2 = \frac{R^2}{4} \left( - \cosh^2 \rho dt^2 +
d\rho^2 + \sinh^2 \rho d\hat{\Omega}^2_2 \right) + R^2 ds_{\C P^3}^2
\end{equation}
where the $\C P^3$ metric is
\begin{equation}
\label{cp3} ds_{\C P^3}^2 = d\theta^2 + \frac{\cos^2 \theta}{4}
d\Omega_2^2 + \frac{\sin^2 \theta}{4} d{\Omega_2'}^2 +  4 \cos^2
\theta \sin^2 \theta ( d \delta + \omega )^2
\end{equation}
with
\begin{equation}
\omega = \frac{1}{4} \sin \theta_1 d\varphi_1 + \frac{1}{4} \sin
\theta_2 d\varphi_2
\end{equation}
Here the curvature radius $R$ is given by
\begin{equation}
R^4 = 32 \pi^2 \lambda l_s^4
\end{equation}
Furthermore, the $\ads_4\times \C P^3$ background has a constant
dilaton with the string coupling given by
\begin{equation}
g_s = \Big( \frac{32\pi^2 \lambda}{k^4} \Big)^{\frac{1}{4}}
\end{equation}
and it has a two-form and a four-form Ramond Ramond flux that will
not be needed here, see for example
\cite{Nishioka:2008gz,Grignani:2008is}. For our purposes it is
convenient to make the coordinate change
\begin{equation}
\psi = 2\theta - \frac{\pi}{2}
\end{equation}
such that the $\C P^3$ metric \eqref{cp3} takes the form
\begin{equation}
\label{cp32} ds_{\C P^3}^2 = \frac{1}{4} d\psi^2 + \frac{1-\sin
\psi}{8} d\Omega_2^2 + \frac{1+\sin \psi}{8} d{\Omega_2'}^2 + \cos^2
\psi ( d \delta + \omega )^2
\end{equation}

The $SU(2)\times SU(2)$ sector corresponds to the two two-spheres in
the $\C P^3$ metric \eqref{cp32}, parameterized as
\begin{equation}
d\Omega_2^2 = d\theta_1^2 + \cos^2 \theta_1 d\varphi_1^2 \spa
{d\Omega_2'}^2 = d\theta_2^2 + \cos^2 \theta_2 d\varphi_2^2
\end{equation}
On the string theory side, the $SU(2)\times SU(2)$ symmetry of the
two two-spheres is a subgroup of the $SU(4)$ symmetry of $\C P^3$.
We can take the three independent Cartan generators for the $SU(4)$
symmetry to be
\begin{equation}
S_z^{(1)} = - i \partial_{\varphi_1} \spa S_z^{(2)} = - i
\partial_{\varphi_2} \spa J = - \frac{i}{2} \partial_\delta
\end{equation}
where $S_z^{(i)}$ are the Cartan generators of the two two-spheres.

On the gauge theory side, the $SU(2)\times SU(2)$ sector corresponds
to consider single-trace operators of the form
\cite{Minahan:2008hf,Gaiotto:2008cg}
\begin{equation}
\label{su2op2} \CO = W^{j_1 j_2 \cdots j_J}_{i_1 i_2 \cdots i_J} \tr
( A_{i_1} B_{j_1} \cdots A_{i_J} B_{j_J} )
\end{equation}
where $A_{1,2}$ and $B_{1,2}$ are the two pairs of complex scalars
in ABJM theory, transforming in the $(1/2,0)$ and $(0,1/2)$ of
$SU(2)\times SU(2)$, respectively, and all scalars being in the
bifundamental representation of $SU(N)\times SU(N)$. Thus, on the
gauge theory side $S_z^{(1)}$ counts the total spin for the
$A_{1,2}$ scalars in \eqref{su2op2} and $S_z^{(2)}$ for the
$B_{1,2}$ scalars. Instead the bare scaling dimension of each scalar
is $1/2$ which means that the total conformal dimension of
\eqref{su2op2} is $\Delta_0 = J$, $\Delta_0$ being the bare scaling
dimension. Indeed, one can define the $SU(2)\times SU(2)$ sector as
consisting of the operators with $\Delta_0 = J$
\cite{Grignani:2008is}.

The energy of the string states in units of the curvature radius $R$
is dual to the scaling dimension $\Delta$ on the gauge theory side.
In terms of the coordinates in the metric \eqref{adscp} we measure
$\Delta$ as
\begin{equation}
\Delta = i \partial_t
\end{equation}

%%%%%%%%%%%%%%%%%%%%%%%%%%%%%%%%%%%%%%%%%%%%%%%%%%%%%%%%%%%%%
\section{Curvature corrections to Penrose limit}
\label{sec:penrose}

In this section we study curvature corrections to the Penrose limit
of \cite{Grignani:2008is}.

\subsection{$SU(2)\times SU(2)$ Penrose limit of $\ads_4\times \C P^3$}

Consider the $\ads_4\times \C P^3$ metric given by \eqref{adscp} and
\eqref{cp32}. We make the coordinate transformation
\begin{equation}
t' = t \spa \chi = \delta - \frac{1}{2} t
\end{equation}
This gives the following metric for $\ads_4\times \C P^3$
\begin{align}
\label{adscp2} ds^2 = & - \frac{R^2}{4} {dt'}^2 ( \sin^2 \psi  +
\sinh^2 \rho
) + \frac{R^2}{4}( d\rho^2 + \sinh^2 \rho d\hat{\Omega}_2^2 ) \nn \\
& + R^2 \left[ \frac{d\psi^2}{4} + \frac{1-\sin \psi}{8} d\Omega_2^2
+ \frac{1+\sin \psi}{8} d{\Omega_2'}^2 + \cos^2 \psi ( dt' + d \chi
+ \omega )( d \chi + \omega )\right]
\end{align}
We have that
\begin{equation}
\label{newch} E \equiv \Delta - J = i \partial_{t'} \spa 2 J = - i
\partial_\chi
\end{equation}

Define the coordinates
\begin{equation}
v = R^2 \chi \spa x_1 = R \varphi_1 \spa y_1 = R \theta_1 \spa x_2 =
R \varphi_2 \spa y_2 = R \theta_2 \spa u_4 = \frac{R}{2} \psi
\end{equation}
We furthermore define $u_1$, $u_2$ and $u_3$ by the relations
\begin{equation}
\frac{R}{2} \sinh \rho = \frac{u}{1 - \frac{u^2}{R^2}} \spa
\frac{R^2}{4} ( d\rho^2 + \sinh^2 \rho d\hat{\Omega}_2^2 ) =
\frac{\sum_{i=1}^3 du_i^2}{(1-\frac{u^2}{R^2}  )^2} \spa u^2 =
\sum_{i=1}^3 u_i^2
\end{equation}
Written explicitly, the metric \eqref{adscp2} in these coordinates
becomes
\begin{align}
\label{adscp3} &ds^2 = -  {dt'}^2 \left( \frac{R^2}{4} \sin^2
\frac{2u_4}{R} + \frac{u^2}{(1 - \frac{u^2}{R^2})^2} \right) +
\frac{\sum_{i=1}^3 du_i^2}{(1-\frac{u^2}{R^2}  )^2} + du_4^2\nn \\
&
+\frac{1}{8}\left(\cos\frac{u_4}{R}-\sin\frac{u_4}{R}\right)^2\left(
dy_1^2+\cos^2\frac{y_1}{R}d
x_1^2\right)+\frac{1}{8}\left(\cos\frac{u_4}{R}+\sin\frac{u_4}{R}\right)^2\left(
dy_2^2+\cos^2\frac{y_2}{R}d x_2^2\right)
\cr&+R^2\cos^2\frac{2u_4}{R} \left[ dt' + \frac{dv}{R^2}
+\frac{1}{4}\left(\sin \frac{y_1}{R} \frac{d x_1}{R} + \sin
\frac{y_2}{R} \frac{d x_2}{R}\right)\right] \left[\frac{dv}{R^2}
+\frac{1}{4}\left(\sin \frac{y_1}{R} \frac{d x_1}{R} + \sin
\frac{y_2}{R} \frac{d x_2}{R}\right)\right]
\end{align}
a very convenient form to expand around $R\to\infty$.

The $SU(2) \times SU(2)$ Penrose limit $R \rightarrow \infty$ of
\cite{Grignani:2008is} gives now the pp-wave metric%
\footnote{See \cite{Bertolini:2002nr} for the analogous Penrose
limit for the $SU(2)$ sector of $\ads_5\times S^5$.}
\begin{equation}
\label{ppwavemetric} ds^2 =  dv dt'  + \sum_{i=1}^4 ( du_i^2 - u_i^2
{dt'}^2 ) + \frac{1}{8} \sum_{i=1}^2 ( dx_i^2 + dy_i^2 + 2 dt' y_i
dx_i )
\end{equation}
The light-cone coordinates in this metric are $t'$ and $v$. We
record here for completeness the two-form and four-form
Ramond-Ramond fluxes
\begin{equation}
\label{ppwaverrfluxes} F_{(2)} = dt' du_4 \spa F_{(4)} = 3 dt' du_1
du_2 du_3
\end{equation}
This is a pp-wave background with 24 supersymmetries first found in
\cite{Sugiyama:2002tf,Hyun:2002wu}. See
\cite{Nishioka:2008gz,Gaiotto:2008cg} for other Penrose limits of
the $\ads_4\times \C P^3$ background giving the pp-wave background
\eqref{ppwavemetric}-\eqref{ppwaverrfluxes}.

We see from \eqref{newch} that
\begin{equation}
\frac{2 J}{R^2} = - i \partial_v
\end{equation}
Thus, the Penrose limit on the gauge theory side is the following
limit
\begin{equation}
\label{penrlim} \lambda, J \rightarrow \infty \ \ \mbox{with}\ \
\lambda' \equiv \frac{\lambda}{J^2} \ \mbox{fixed} \spa \Delta-J \
\mbox{fixed}
\end{equation}

\subsection{Bosonic string Hamiltonian}

We now consider type IIA string theory on $\ads_4\times \C P^3$ in
the above Penrose limit, including the curvature corrections in
$1/R$. For simplicity we consider only the bosonic string modes. We
set the string length $l_s =1$ in the rest of this paper.

The bosonic string action is given by
\begin{equation}
I = \frac{1}{2\pi } \int d\tau d\sigma \CL \spa \CL = - \frac{1}{2}
h^{\alpha \beta} G_{\mu\nu}
\partial_\alpha x^\mu
\partial_\beta x^\nu
\end{equation}
Here $h^{\alpha \beta} = \sqrt{-\det \gamma} \gamma^{\alpha \beta}$
with $\gamma_{\alpha \beta}$ being the world-sheet metric. This
means that $\det h = -1$, thus $h^{\alpha\beta}$ has only two
independent components. The metric $G_{\mu\nu}$ is given by
\eqref{adscp3}.

For convenience we define the momenta as
\begin{equation} \label{ps}
p_\mu = - h^{\tau \alpha} G_{\mu \nu} \partial_\alpha x^\nu
\end{equation}
From this we see that
\begin{equation} \label{dots}
\dot{x}^\mu = - \frac{1}{h^{\tau\tau}} G^{\mu\nu} p_\nu -
\frac{h^{\tau\sigma}}{h^{\tau\tau}} {x'}^\mu
\end{equation}
\begin{equation}
\CL = - \frac{1}{2 h^{\tau\tau}} G^{\mu\nu} p_\mu p_\nu + \frac{1}{2
h^{\tau\tau}} G_{\mu\nu} {x'}^\mu {x'}^\nu
\end{equation}
The Hamiltonian density is
\begin{equation}
\CH = p_\mu \dot{x}^\mu - \CL = - \frac{1}{2h^{\tau\tau}} \left(
G^{\mu\nu} p_\mu p_\nu + G_{\mu\nu} {x'}^\mu {x'}^\nu \right) -
\frac{h^{\tau\sigma}}{h^{\tau\tau}} {x'}^\mu p_\mu
\end{equation}
Considering the two fields $\frac{1}{h^{\tau\tau}}$ and
$\frac{h^{\tau\sigma}}{h^{\tau\tau}}$ as the two independent
components of $h^{\alpha\beta}$, we can regard these two fields as
Lagrange multipliers. This gives the constraints
\begin{equation}
\label{constr1} G^{\mu\nu} p_\mu p_\nu + G_{\mu\nu} {x'}^\mu
{x'}^\nu = 0 \spa {x'}^\mu p_\mu = 0
\end{equation}

We impose now the lightcone gauge
\begin{equation}
t' = c \tau \spa p_v = \mbox{const.}
\end{equation}
where $c$ is a constant. The constant $c$ can be fixed from the term
$\frac{c}{2}\partial_{\tau}v$ in the full Lagrangian. In fact we
have that $ p_v=\partial \CL/\partial \partial_{\tau} v$ which gives
\begin{equation}
c=\frac{4J}{R^2}=\frac{J}{\pi\sqrt{2\lambda}} \label{cconstant}
\end{equation}
where we used that $\int_0^{2\pi}\frac{d\sigma}{2\pi} p_\chi=2J$.
Then the constraints \eqref{constr1} can be written as
\begin{eqnarray}
\label{constr2a} && G^{t't'} (p_{t'})^2+G^{vv} (p_v)^2 + 2 G^{t'v}
p_{t'} p_v + 2 G^{t' x_a} p_{t'} p_{x_a} + 2 G^{vx_a} p_v p_{x_a} +
G^{x_a x_b} p_{x_a} p_{x_b}+G^{y_a y_a} p_{y_a} p_{y_a} \cr
&&+G^{u_i u_j}p_{u_i} p_{u_j}+G_{vv}\left(v'\right)^2+ 2 G_{v x_a}
{v}' {x_a}' + G_{x_a x_b} {x_a'} {x_b'}+G_{y_a y_a} {y_a'} {y_a'} +
G_{u_i u_j} {u_i'} {u_j'} = 0
\end{eqnarray}
\begin{equation}
\label{constr2b} {v'} p_v + {x_a}'p_{x_a}+{y_a}'p_{y_a}+ {u_i'}
p_{u_i} = 0
\end{equation}
with $a,b=1,2$ and $i,j=1,2,3,4$. Eliminating $v'$ in
\eqref{constr2a} using \eqref{constr2b}, one gets a quadratic
equation for the light-cone Hamiltonian density
 $\CH^{\rm lc} = - p_{t'}$. Thus, we can solve the quadratic constraint
\eqref{constr2a} to obtain the lightcone Hamiltonian density, which
we then expand up to $\CO \left(\frac{1}{R^2}\right)$
\begin{equation}
\label{genhlc} \CH^{\rm lc} =\CH^{\rm lc}_{\rm free}+\CH^{\rm
lc}_{\rm int}
\end{equation}
The complete expression for the Hamiltonian $\CH^{\rm lc}_{\rm int}$
in terms of the momenta is however quite complicated even at the
order $\CO \left(\frac{1}{R^2}\right)$. So we do not reproduce it
here. It simplifies a lot instead when written in terms of the
velocities at the zeroth order in the $\frac{1}{R}$ expansion.

To eliminate the momenta in terms of the velocities we should use
eq.(\ref{ps}) with the leading order worldsheet metric $h^{\tau
\tau}=-1$, $h^{\tau \sigma}=0$. One gets
\begin{equation} \label{pscallan}
p_{x_1}=\frac{1}{8}\left(\dot x_1+c y_1\right)  \spa
p_{x_2}=\frac{1}{8}\left(\dot x_2+c y_2\right)\spa
p_{y_1}=\frac{1}{8}\dot y_1\spa p_{y_2}=\frac{1}{8}\dot y_2
\end{equation}
where by $\dot x_a,~\dot y_a$ we mean the velocities at the zeroth
order in the $\frac{1}{R}$ expansion. The other momenta are
standard. The leading term in the $\frac{1}{R}$ expansion gives the
pp-wave quadratic Hamiltonian
\begin{align} \label{hfreecallan2} \CH^{\rm lc}_{\rm
free}=\frac{1}{16
c}\left[\left(x'_a\right)^2+\left(y'_a\right)^2+\left(\dot{x}_a\right)^2+
\left(\dot{y}^2_a\right)^2\right] + \frac{1}{2c} \sum_{i=1}^4 \Big[
(\dot u_i)^2 +(u_i')^2 + c^2 u_i^2 \Big]
\end{align}
The interacting Hamiltonian contains two parts, one that goes like
$1/R$ which is cubic in the fields and the other one that goes like
$1/R^2$ which is quartic in the fields
\begin{equation}
\CH^{\rm lc}_{\rm int}=\CH^{(1)}_{\rm int}+\CH^{(2)}_{\rm int}
\end{equation}
where
\begin{equation}\label{ch1}
\CH^{(1)}_{\rm int}=\frac{u_4}{8 R c}\left[(\dot x_1)^2-(\dot
x_2)^2+(\dot y_1)^2-(\dot
y_2)^2-(x_1')^2+(x_2')^2-(y_1')^2+(y_2')^2\right]
\end{equation}
and
\begin{align}\label{ch2}
&\CH^{(2)}_{\rm int}=\frac{1}{128 R^2 c^3}\left[4\left(\dot x_a
x_a'+\dot y_a
y_a'\right)^2-\left(\left(x'_a\right)^2+\left(y'_a\right)^2+\left(\dot{x}_a\right)^2+
\left(\dot{y}_a\right)^2\right)^2\right]\cr&+\frac{1}{48 R^2
c}\left[3  \left(\left((\dot x_1
)^2-(x'_1)^2\right)y_1^2+\left((\dot x_2
)^2-(x'_2)^2\right)y_2^2\right)+c\left(\dot x_1 y_1^3+\dot x_2
y_2^3\right)\right]+\dots
\end{align}
the dots are for terms that are irrelevant in the computation of the
spectrum of string states belonging to the  $SU(2)\times SU(2)$
sector.

From the Hamiltonian densities one gets the Hamiltonian as
\begin{equation}
H_{\rm free}=\frac{1}{2\pi}\int_{0}^{2\pi}\CH^{\rm lc}_{\rm free}
d\sigma\spa H_{\rm int}=\frac{1}{2\pi}\int_{0}^{2\pi}\CH^{\rm
lc}_{\rm int} d\sigma
\end{equation}
The mode expansion for the bosonic fields can be written as
\begin{equation}
u_i (\tau,\sigma ) = i \frac{1}{\sqrt{2}} \sum_{n\in \Z}
\frac{1}{\sqrt{\Omega_n}} \Big[ \hat{a}^i_n e^{-i ( \Omega_n \tau -
n \sigma ) } - (\hat{a}^i_n)^\dagger e^{i ( \Omega_n \tau - n \sigma
) } \Big]
\end{equation}
\begin{equation}
\label{zmode} z_a(\tau,\sigma) = 2 \sqrt{2} \, e^{i\frac{ c
\tau}{2}} \sum_{n \in \Z} \frac{1}{\sqrt{\omega_n}} \Big[ a_n^a
e^{-i ( \omega_n \tau - n \sigma ) } -  (\tilde{a}^a)^\dagger_n e^{i
( \omega_n \tau - n \sigma ) } \Big]
\end{equation}
where $\Omega_n=\sqrt{c^2+n^2}$, $\omega_n=\sqrt{\frac{c^2}{4}+n^2}$
and we defined
$z_a(\tau,\sigma)=x_a(\tau,\sigma)+iy_a(\tau,\sigma)$. The canonical
commutation relations $[x_a(\tau,\sigma),p_{x_b}(\tau,\sigma')] =
i\delta_{ab} \delta (\sigma-\sigma')$,
$[y_a(\tau,\sigma),p_{y_b}(\tau,\sigma')] = i\delta_{ab}\delta
(\sigma-\sigma')$ and $[u_i(\tau,\sigma),p_j(\tau,\sigma')] =
i\delta_{ij} \delta (\sigma-\sigma')$ follow from
\begin{equation}
\label{comrel} [a_m^a,(a_n^b)^\dagger] = \delta_{mn} \delta_{ab}\spa
[\tilde{a}_m^a,(\tilde{a}_n^b)^\dagger] = \delta_{mn}
\delta_{ab}\spa [\hat{a}^i_m,(\hat{a}^j_n)^\dagger] = \delta_{mn}
\delta_{ij}
\end{equation}
Employing \eqref{comrel} and \eqref{hfreecallan2} we obtain the
bosonic free Hamiltonian as
\begin{equation}
c H_{\rm free} = \sum_{i=1}^4 \sum_{n\in \Z} \sqrt{n^2+c^2}\,
\hat{N}^i_n+\sum_{a=1}^2\sum_{n\in \Z}
\left(\sqrt{\frac{c^2}{4}+n^2} - \frac{c}{2}\right) M_n^a
+\sum_{a=1}^2\sum_{n\in \Z} \left(\sqrt{\frac{c^2}{4}+n^2}+
\frac{c}{2}\right) N_n^a \label{penspectrum}
\end{equation}
with the number operators $\hat{N}^i_n = (\hat{a}^i_n)^\dagger
\hat{a}^i_n$, $M_n^a = (a^a)^\dagger_n a^a_n$ and $N_n^a =
(\tilde{a}^a)^\dagger_n \tilde{a}_n^a$, and with the level-matching
condition
\begin{equation}
\label{levelm} \sum_{n\in \Z}n \left[\sum_{i=1}^4
\hat{N}^i_n+\sum_{a=1}^2 \left(M_n^a + N_n^a\right)\right]
 = 0
\end{equation}
Using \eqref{cconstant} the spectrum \eqref{penspectrum} reads
\begin{equation}\small{
 H_{\rm free} = \sum_{i=1}^4 \sum_{n\in \Z}
\sqrt{1+\frac{2\pi^2\lambda}{J^2}n^2}
\hat{N}^i_n+\sum_{a=1}^2\sum_{n\in \Z}\left[
\left(\sqrt{\frac{1}{4}+\frac{2\pi^2\lambda}{J^2}n^2} -
\frac{1}{2}\right) M_n^a +
\left(\sqrt{\frac{1}{4}+\frac{2\pi^2\lambda}{J^2}n^2}+
\frac{1}{2}\right) N_n^a\right]} \label{penH}
\end{equation}
\subsection{Perturbative analysis of the string energy spectrum}

We shall now compute finite size corrections to the energies of two
oscillator states of the form
\begin{equation}\label{s}
|s\rangle = (a_n^1)^\dagger (a_{-n}^1)^\dagger|0\rangle
\end{equation}
with both oscillators in just one of the two $SU(2)$'s of the
$SU(2)\times SU(2)$ sector, and of the form
\begin{equation}\label{t}
|t\rangle = (a^1_n)^\dagger (a^2_{-n})^\dagger|0\rangle
\end{equation}
with an oscillator in each of the two $SU(2)$'s of the $SU(2)\times
SU(2)$ sector.

At the first order in perturbation theory the Hamiltonian
\eqref{ch1} does not contribute to the energies of the states
\eqref{s} and \eqref{t}. Its mean value on these states vanishes, so
that we shall only have corrections to the energies at the order
$\CO\left(\frac{1}{R^2}\right)$. We will thus have two contributions
to the energy corrections, one that comes from computing at the
second perturbative order the contribution of the term \eqref{ch1}
and one that arises from the first perturbative order just by taking
the mean value of the Hamiltonian \eqref{ch2} on the states
$|s\rangle$ and $|t\rangle$. For these states we have respectively
\begin{equation}\label{energycorrection}
E_{s,t}^{(2)}=\langle s, t|H^{(2)}_{\rm
int}|s,t\rangle+\sum_{|i\rangle}\frac{\left|\langle i|H^{(1)}_{\rm
int}|s,t\rangle\right|^2}{E^{(0)}_{|s\rangle,|t\rangle}-E^{(0)}_{|i\rangle}}
\end{equation}
where $|i\rangle$ is an intermediate state with zeroth order energy
$E^{(0)}_{|i\rangle}$.

 The relevant part of the Hamiltonian \eqref{ch1} contributing to the second term in
 \eqref{energycorrection} written in terms of oscillators reads
\begin{equation}\label{hc1osc}
H_{\rm int}^{(1)}=\frac{i}{Rc\sqrt{2}}\sum_{m,\,l,\,r}
\frac{1}{\sqrt{\omega_m\omega_l\Omega_r}}
\left[\left(\omega_m-\frac{c}{2}\right)\left(\omega_l-\frac{c}{2}\right)+ml\right]
(\hat a_{-r}^4)^\dagger\left[(a_{-m}^2)^\dagger
(a_l^2)-(a_{-m}^1)^\dagger (a_l^1)\right]
\end{equation}
We have written operator monomials in normal ordered form. The
normal ordering ambiguity that would arise in $H_{int}^{(1)}$  would
not contribute to the matrix elements in \eqref{energycorrection}.

The quartic part of the interaction Hamiltonian \eqref{ch2} which is
relevant for computing the $\CO \left(\frac{1}{R^2}\right)$
corrections to the pp-wave spectrum reads
\begin{eqnarray} \label{hamoscillators}
&& H^{(2)}_{\rm int}  = \sum_{m,p,q,r}
\frac{\left[(a_{-m}^1)^\dagger (a_{-p}^1)^\dagger a^1_q
a^1_r+(a_{-m}^2)^\dagger (a_{-p}^2)^\dagger a^2_q
a^2_r\right]\delta\left(m+p+q+r\right)}{R^2\sqrt{\omega_m \omega_p
\omega_q \omega_r}}\cr &&\Bigg\{\frac{-m p-q r+4 m q}{4
c}+\frac{\omega_m+\omega_q-c}{4}-\frac{m p q r}{2 c^3}\frac{1}{4
c}\left[\left(\omega_m-\frac{c}{2}\right)\left(\omega_p-\frac{c}{2}\right)+
\left(\omega_q-\frac{c}{2}\right)\left(\omega_r-\frac{c}{2}\right)\right.\cr
&&\left.+4
\left(\omega_m-\frac{c}{2}\right)\left(\omega_q-\frac{c}{2}\right)\right]+\frac{1}{2
c^3}\left[m p
\left(\omega_q-\frac{c}{2}\right)\left(\omega_r-\frac{c}{2}\right)+q
r
\left(\omega_m-\frac{c}{2}\right)\left(\omega_p-\frac{c}{2}\right)\right]\cr
&&-\frac{1}{2
c^3}\left(\omega_m-\frac{c}{2}\right)\left(\omega_p-\frac{c}{2}\right)\left(\omega_q-\frac{c}{2}\right)
\left(\omega_r-\frac{c}{2}\right)\Bigg\}\cr && - \sum_{m,p,q,r}
\frac{(a_{-m}^1)^\dagger (a_{p}^1) (a_{-q}^2)^\dagger
a^2_r\delta\left(m+p+q+r\right)}{R^2 c^3\sqrt{\omega_m \omega_p
\omega_q \omega_r}}\Bigg\{\left[\left(\omega_m-\frac{c}{2}\right)
\left(\omega_p-\frac{c}{2}\right)-m p\right]\times \cr
&&\left[\left(\omega_q-\frac{c}{2}\right)\left(\omega_r-\frac{c}{2}\right)-q
r\right]-\left[m\left(\omega_p-\frac{c}{2}\right)-p\left(\omega_m-\frac{c}{2}\right)\right]
\left[q\left(\omega_r-\frac{c}{2}\right)-r\left(\omega_q-\frac{c}{2}\right)\right]\Bigg\}\cr
&&
\end{eqnarray}
Also in this case we have chosen to write operators in normal
ordered form. Since $H_{int}^{(2)}$ was derived as a classical
object, it does not follow what the correct ordering of the
operators is. A non-zero normal ordering constant would give a
contribution to the Hamiltonian of the form
\begin{equation}
H_{\rm norm.ord.} = \sum_n C_n \Big( (a_n^1)^\dagger a_n^1 +
(a_n^2)^\dagger a_n^2 \Big)
\end{equation}
We assume in this paper that $C_n=0$. Presumably one can argue for
this on the same lines as in \cite{Callan:2003xr,Callan:2004uv}.
Moreover, one can consider the single-magnon state $(a_n^1)^\dagger
|0\rangle$ which, based on the general dispersion relation
\eqref{dispgen}, should not receive $1/J$ corrections. This is
consistent with $C_n=0$. Finally, we shall see in Section
\ref{sec:compare} that we get agreement for the $|s\rangle$ and
$|t\rangle$ string states with the Bethe ansatz assuming $C_n=0$.

We now compute the energies of the states $|s\rangle$ and
$|t\rangle$ \eqref{s}-\eqref{t}. Consider first the state $|t\rangle
=(a_n^1)^\dagger (a^2_{-n})^\dagger|0\rangle$. To derive the mean
value of \eqref{hamoscillators} we need the following quantity
$$
\langle 0| (a_n^1) (a^2_{-n})(a^1_{-m})^\dagger a^1_p
(a^2_{-q})^\dagger a^2_r (a_n^1)^\dagger (a^2_{-n})^\dagger|0\rangle
= \delta_{m,-n} \delta_{p,n}\delta_{q,n} \delta_{r,-m}
$$
so that the mean value of \eqref{hamoscillators} contributing to
\eqref{energycorrection} reads
\begin{equation} \label{mixingspectrum1}
\langle t| H^{(2)}_{\rm int}|t\rangle
=-\frac{\left[n^2+\left(\omega_n-\frac{c}{2}\right)^2\right]^2+
4n^2\left(\omega_n-\frac{c}{2}\right)^2}{R^2 c^3 \omega_n^2}\simeq
-\frac{4 n^4 \pi^4 \lambda'^2}{J} +\frac{16 n^6 \pi^6
\lambda'^3}{J}+\CO\left(\lambda'^4\right)
\end{equation}
where $\lambda'$ is defined in \eqref{penrlim} and we used that
$R^2=4\pi\sqrt{2\lambda}$.

To compute the second term in
\eqref{energycorrection} we need to consider intermediate states
that
 give a non vanishing matrix element for the $H^{(1)}_{\rm int}$ given in \eqref{hc1osc}. The only
possible intermediate states that have this property are three
oscillator states of the form $(a^4_{-p-q})^\dagger
(a^1_{p})^\dagger(a^2_{q})^\dagger|0\rangle$. Computing the matrix
element is simple and we get
\begin{equation}\label{mixingspectrum2}
\sum_{|i\rangle}\frac{\left|\langle i|H^{(1)}_{\rm
int}|t\rangle\right|^2}{E^{(0)}_{|t\rangle}-E^{(0)}_{|i\rangle}}=\frac{1}{
R^2c}\sum_p\frac{\left[\left(\omega_{p+n}-\frac{c}{2}\right)\left(\omega_{n}-\frac{c}{2}\right)-(p+n)n\right]^2}
{\omega_{p+n}\omega_n\Omega_p\left(\omega_{p+n}-\omega_n-\Omega_p\right)}
+\frac{\left[\left(\omega_{n}-\frac{c}{2}\right)^2-n^2\right]^2}{R^2
c^3\omega^2_{n}}
\end{equation}
Using $\zeta$-function regularization the first term vanishes, so
that for the $\CO\left(\frac{1}{R^2}\right)$ correction to the
energy of the state $|t\rangle$, we get, adding
\eqref{mixingspectrum1} and \eqref{mixingspectrum2}
\begin{equation}\label{Et}
E_t^{(2)}=-\frac{\left[n^2+\left(\omega_n-\frac{c}{2}\right)^2\right]^2+
4n^2\left(\omega_n-\frac{c}{2}\right)^2}{R^2 c^3
\omega_n^2}+\frac{\left[\left(\omega_{n}-\frac{c}{2}\right)^2-n^2\right]^2}{
R^2c^3\omega^2_{n}}\simeq -\frac{64 n^6
\pi^6\lambda'^3}{J}+\CO\left(\lambda'^4\right)
\end{equation}
It is interesting to note that for the state $|t\rangle$ the first
finite-size correction appears at the order $\lambda'^3$. In
particular, that the finite-size correction is zero at order
$\lambda'^2$ is due to a rather non-trivial cancelation of the
mean-value contribution of \eqref{ch2} and the contribution coming
from the $1/\sqrt{J}$ interaction term \eqref{ch1}, which enters
through a second-order perturbative energy correction and is
regularized using $\zeta$-function regularization.

Consider now the state $|s\rangle = (a_n^1)^\dagger
(a_{-n}^1)^\dagger|0\rangle$. Since we have that
$$
\langle 0| a^1_n a^1_{-n} (a^1_{-m})^\dagger a^1_p
(a^2_{-q})^\dagger a^2_r (a^1_n)^\dagger (a^1_{-n})^\dagger|0\rangle
= \left(\delta_{m,n}\delta_{p,-n}+\delta_{m,-n}\delta_{p,n}\right)
\left(\delta_{q,n}\delta_{r,-n}+\delta_{r,n}\delta_{q,-n}\right)
$$
one gets
\begin{eqnarray} \label{samesu2spectrum1}
&&\langle s| H^{(2)}_{\rm int}
|s\rangle=-\frac{2\left[\left(\omega_n-c\right)\left(4n^2-c^2\right)-c^2
\omega_n\right]}{R^2 c^3 \omega_n}\simeq \frac{8 n^2 \pi^2
\lambda'}{J}-\frac{56 n^4 \pi^4 \lambda'^2}{J}+\frac{352 n^6
\pi^6\lambda'^3}{J}+\CO\left(\lambda'^4\right)\cr&&
\end{eqnarray}
To compute the second term in \eqref{energycorrection} we need to
consider intermediate states of the form $(a^4_{-p-q})^\dagger
(a^1_{p})^\dagger(a^1_{q})^\dagger|0\rangle$. Computing the matrix
element of \eqref{hc1osc}, the second term in
\eqref{energycorrection} gives the contribution
\begin{eqnarray}\label{samesu2spectrum2}
\sum_{|i\rangle}\frac{\left|\langle i|H^{(1)}_{\rm
int}|s\rangle\right|^2}{E^{(0)}_{|s\rangle}-E^{(0)}_{|i\rangle}}&&=\frac{1}{
R^2c}\sum_p\frac{\left[\left(\omega_{p+n}-\frac{c}{2}\right)\left(\omega_{n}-\frac{c}{2}\right)-(p+n)n\right]^2}
{\omega_{p+n}\omega_n\Omega_p\left(\omega_{p+n}-\omega_n-\Omega_p\right)}\cr&&-
\frac{\left[\left(\omega_{n}-\frac{c}{2}\right)^2+n^2\right]^2}{R^2
c\,\omega^2_{n}\Omega^2_{2n}}
-\frac{\left[\left(\omega_{n}-\frac{c}{2}\right)^2-n^2\right]^2}{R^2
c^3\omega^2_{n}}
\end{eqnarray}
where we have divided by 2 to avoid overcounting of intermediate
states. Using $\zeta$-function regularization the first term
vanishes, so that for the $\CO\left(\frac{1}{R^2}\right)$ correction
to the energy of the state $|s\rangle$, adding
\eqref{samesu2spectrum1} and \eqref{samesu2spectrum2}, we get
\begin{eqnarray}\label{Es}
E_s^{(2)}&=&-2\frac{\left[\left(\omega_n-c\right)\left(4n^2-c^2\right)-c^2
\omega_n\right]}{R^2 c^3 \omega_n}-
\frac{\left[\left(\omega_{n}-\frac{c}{2}\right)^2+n^2\right]^2}{R^2
c\,\omega^2_{n}\Omega^2_{2n}}
-\frac{\left[\left(\omega_{n}-\frac{c}{2}\right)^2-n^2\right]^2}{R^2
c^3\omega^2_{n}}\cr&\simeq &\frac{8 n^2 \pi^2 \lambda'}{J}-\frac{64
n^4 \pi^4 \lambda'^2}{J}+\frac{448 n^6
\pi^6\lambda'^3}{J}+\CO\left(\lambda'^4\right)
\end{eqnarray}
%

%%%%%%%%%%%%%%%%%%%%%%%%%%%%%%%%%%%%%%%%%%%%%%%%%%%%%%%%%%%%%
\section{Low energy sigma-model expansion}
\label{sec:smexp}

In this section we consider the low energy sigma-model expansion in
which $\Delta -J$ is small. In this way we zoom in to the
$SU(2)\times SU(2)$ sector on the string side with $\lambda' =
\lambda / J^2$ being small. To leading order we reproduce the result
of \cite{Grignani:2008is} that the sigma-model consists of two
Landau-Lifshitz models added together without interaction. We then
move on to obtain the first and second order corrections in
$\lambda'$ to the leading sigma-model. We compare the energies of
the $|s\rangle$ and $|t\rangle$ string states found in Section
\ref{sec:penrose} for the $\lambda'$, ${\lambda'}^2$ and
${\lambda'}^3$ orders and find agreement.

The methods that we employ in this section have been developed in
\cite{Kruczenski:2003gt,Kruczenski:2004kw,Minahan:2005mx,Astolfi:2008yw,Harmark:2008gm}.

\subsection{Expansion of sigma-model action}

We want to extract the effective sigma-model description of the
$SU(2)\times SU(2)$ sector, including corrections in $\lambda'$.
Define
\begin{equation}
\label{xpm} x^+ = \lambda' t \spa x^- = \delta - \frac{1}{2} t
\end{equation}
with
\begin{equation}
\lambda' \equiv \frac{\lambda}{J^2}
\end{equation}
Then the charges are
\begin{equation}
\frac{E}{\lambda'} = \frac{ \Delta - J }{\lambda'} = - P_+ = i
\partial_{x^+} \spa P_- = - i \partial_{x^+} = 2J
\end{equation}
We see that taking the $\lambda' \rightarrow 0$ limit means that
$\Delta - J\rightarrow 0$ which means that we keep the modes of the
$SU(2)\times SU(2)$ sector dynamical, while the other modes become
non-dynamical in this limit. Naively, this leads to the reasoning
that one can set $\rho=0$ and $\psi=0$ in the $\ads_4\times \C P^3$
background \eqref{adscp} with the $\C P^3$ metric given by
\eqref{cp32}. However, as we shall see in the following the field
$\psi$ does couple to the modes of the $SU(2)\times SU(2)$ sector
even though it becomes non-dynamical in the $\lambda' \rightarrow 0$
limit.

Consider therefore the $\ads_4\times \C P^3$ metric given by
\eqref{adscp} and \eqref{cp32} with $\rho=0$ and in terms of the
$x^+$, $x^-$ coordinates \eqref{xpm}
\begin{align}
\label{startmet} ds^2 = & R^2 \Big[ - \frac{1}{4{\lambda'}^2} \sin^2
\psi (dx^+)^2 + \frac{1}{4} d\psi^2 + \cos^2 \psi ( {\lambda'}^{-1}
dx^+ + dx^- + \omega ) ( dx^- + \omega ) \nn \\ & + \frac{1-\sin
\psi}{8} d\Omega_2^2 + \frac{1+ \sin \psi}{8} d{\Omega_2'}^2 \Big]
\end{align}
 The
idea in the following is that $\psi$ as expected is non-dynamical in
the $\lambda'\rightarrow 0$ limit, however, one has to include it.
We show that in the $\lambda'\rightarrow 0$ limit $\psi$ acts as a
Lagrange multiplier, and solving the constraint associated to $\psi$
gives extra terms to the effective sigma-model.

We consider the bosonic sigma-model Lagrangian
\begin{equation}
\CL = - \frac{1}{2} h^{\alpha\beta} G_{\mu\nu} \partial_\alpha x^\mu
\partial_\beta x^\nu
\end{equation}
with the Virasoro constraints
\begin{equation}
G_{\mu\nu} ( \partial_\alpha x^\mu \partial_\beta x^\nu -
\frac{1}{2} h_{\alpha\beta} h^{\gamma\delta} \partial_\gamma x^\mu
\partial_\delta x^\nu ) = 0
\end{equation}
with $G_{\mu\nu}$ being the metric \eqref{startmet}. Define for
convenience
\begin{equation}
A \equiv - h^{00} \spa B \equiv h^{01}
\end{equation}
Since the determinant of $h^{\alpha\beta}$ is $-1$ we have $h^{11} =
( 1 - B^2 ) / A $. For $\lambda' \rightarrow 0$ we have that $A = 1$
and $B=0$. Define
\begin{equation}
S_{\alpha\beta} \equiv G_{\mu\nu} \partial_\alpha x^\mu
\partial_\beta x^\nu
\end{equation}
We can now write the Lagrangian as
\begin{equation}
\CL = \frac{A}{2} S_{00} - B S_{01} - \frac{1-B^2}{2A} S_{11}
\end{equation}
and the Virasoro constraints as
\begin{equation}
\label{vircon} \begin{array}{c} \ds (1+B^2) S_{00} +
\frac{2B(1-B^2)}{A} S_{01} + \frac{(1-B^2)^2}{A^2} S_{11} = 0
\\[4mm] \ds
AB S_{00} + 2(1-B^2) S_{01} - \frac{B(1-B^2)}{A} S_{11} =0
\end{array}
\end{equation}

Our gauge choice is
\begin{equation}
x^+ = \kappa \tau
\end{equation}
\begin{equation}
\label{gaugecon} 2\pi  p_- = \frac{\partial \CL}{\partial
\partial_\tau x^-} = \mbox{const.} \spa \frac{\partial \CL}{\partial
\partial_\sigma x^-} = 0
\end{equation}
Thus, we are not fixing the world-sheet metric in this gauge choice,
but rather that the angular momentum $J$ is evenly distributed along
the string \cite{Kruczenski:2004kw}. We have
\begin{equation}
\label{pminus} 2\pi  p_- = R^2 \cos^2 \psi \Big[ \frac{A
\kappa}{2\lambda'} + A (\partial_\tau x^- + \omega_\tau ) - B (
{x^-}' + \omega_\sigma ) \Big]
\end{equation}
The $\psi$ field will be seen to be a non-dynamical field, thus it
should be considered here as a Lagrange-multiplier. For $\lambda'
\rightarrow 0$ we require that $\psi \rightarrow 0$. The dominating
term for $\lambda' \rightarrow 0$ is therefore
\begin{equation}
\label{pminus2} 2\pi  p_- = \frac{R^2 \kappa}{2\lambda'}
\end{equation}
From this we obtain
\begin{equation}
2 J = P_- = \int_0^{2\pi} d\sigma p_- = \frac{R^2 \kappa}{2
\lambda'} = \frac{2\pi \sqrt{2\lambda} \kappa}{\lambda'}
\end{equation}
where we used that $R^2 = 4\pi \sqrt{2\lambda} $. We see from this
that
\begin{equation}
\kappa = \frac{\sqrt{\lambda'}}{\pi \sqrt{2}}
\end{equation}
Thus $\kappa$ goes like $\sqrt{\lambda'}$. This means that $\kappa
\rightarrow 0$ for $\lambda' \rightarrow 0$. Write now
\begin{equation}
\partial_\tau x^\mu = \kappa \dot{x}^\mu
\end{equation}
Then we should keep fixed $\dot{x}^\mu$ in the $\kappa \rightarrow
0$ limit, since that corresponds to the correct energy scale. Define
therefore the rescaled world-sheet time $\tilde{\tau}$ as
$\tilde{\tau} = \kappa \tau$ so that we have $\dot{x}^\mu =
\partial_{\tilde{\tau}} x^\mu$.

Using the metric \eqref{startmet} we compute
\begin{align}
\label{S00} S_{00} = & R^2\kappa^2 \Big[ - \frac{1}{4{\lambda'}^2}
\sin^2 \psi + \frac{1}{4} \dot{\psi}^2 + \cos^2\psi \Big(
\frac{1}{\lambda'} + \dot{x}^- + \dot{\omega} \Big) (\dot{x}^- +
\dot{\omega}) \nn \\ & + \frac{1-\sin \psi}{8} ( \dot{\theta}_1^2 +
\cos^2 \theta_1 \dot{\varphi}_1^2 ) + \frac{1-\sin \psi}{8} (
\dot{\theta}_2^2 + \cos^2 \theta_2 \dot{\varphi}_2^2 ) \Big]
\end{align}
\begin{align}
S_{01} = & R^2\kappa \Big[ \frac{1}{4} \dot{\psi} \psi' +
\cos^2\psi\Big( \frac{1}{2\lambda'} + \dot{x}^- + \dot{\omega} \Big)
( {x^-}' + \omega_\sigma ) \nn \\ &  + \frac{1-\sin \psi}{8} (
\dot{\theta}_1 \theta_1' + \cos^2 \theta_1 \dot{\varphi}_1
\varphi_i' ) + \frac{1+\sin \psi}{8} ( \dot{\theta}_2 \theta_2' +
\cos^2 \theta_2 \dot{\varphi}_2 \varphi_2' ) \Big]
\end{align}
\begin{align}
\label{S11} S_{11} = R^2 \Big[ \frac{1}{4} {\psi'}^2 + \cos^2\psi (
{x^-}' + \omega_\sigma )^2  + \frac{1-\sin\psi}{8} ( {\theta_1'}^2 +
\cos^2 \theta_1 {\varphi_1'}^2 )   + \frac{1+\sin\psi}{8} (
{\theta_2'}^2 + \cos^2 \theta_2 {\varphi_2'}^2 ) \Big]
\end{align}

To find the effective action we should solve the two Virasoro
contraints \eqref{vircon} and the two gauge conditions
\eqref{gaugecon} (with $S_{00}$, $S_{01}$ and $S_{11}$ as in
Eqs.~\eqref{S00}-\eqref{S11}) to obtain $\dot{x}^-$, ${x^-}'$, $A$
and $B$ in terms of the transverse fields and their derivatives.
This we do order by order in $\kappa$. A convenient way to do this
is to first solve the two gauge conditions \eqref{gaugecon} to find
$\dot{x}^-$ and ${x^-}'$ in terms of $A$, $B$ and the transverse
fields. This gives
\begin{equation}
\label{xminussol} \dot{x}^- = - \dot{\omega} - \frac{1}{2\lambda'} +
\frac{1-B^2}{2A\lambda' \cos^2 \psi} \spa {x^-}' = - \omega_\sigma -
\frac{B\kappa}{2\lambda' \cos^2 \psi}
\end{equation}
We subsequently plug this into the Virasoro constraints
\eqref{vircon} to solve for $A$ and $B$ in terms of the transverse
fields and their derivatives. To this end we expand $A$ and $B$ as
follows
\begin{equation}
A = 1 + \kappa^2 A_1 + \kappa^4 A_2 +\cdots \spa B = \kappa^3 B_1 +
\kappa^5 B_2 + \cdots
\end{equation}
We furthermore make the following expansion of $\psi$
\begin{equation}
\psi = \kappa^2 \psi_1 + \kappa^4 \psi_2 + \cdots
\end{equation}

We now solve the Virasoro constraints \eqref{vircon} order by order
in $\kappa$. We get
\begin{equation}
\label{theA1} A_1 = \pi^4 \sum_{i=1}^2 (\vec{n}_i')^2 \spa B_1 = 2
\pi^4 \sum_{i=1}^2 \dot{\vec{n}}_i \cdot \vec{n}_i'
\end{equation}
\begin{align}
A_2 = \frac{\psi_1^2}{2} - \pi^4 \psi_1 [ (\vec{n}_1')^2 -
(\vec{n}_2')^2 ] + \pi^4 [ (\dot{\vec{n}}_1)^2 +
(\dot{\vec{n}}_2)^2] - \frac{\pi^8}{2} [ (\vec{n}_1')^2 +
(\vec{n}_2')^2 ]^2
\end{align}
\begin{align}
\label{theB2} B_2 = - 2 \pi^4 \psi_1 [ \dot{\vec{n}}_1 \cdot
\vec{n}_1' - \dot{\vec{n}}_2 \cdot \vec{n}_2' ] - 2 \pi^8
[(\vec{n}_1')^2 + (\vec{n}_2')^2] [\dot{\vec{n}}_1 \cdot \vec{n}_1'
+ \dot{\vec{n}}_2 \cdot \vec{n}_2']
\end{align}
where we here and in the following simplify our expressions by using
the two unit vector fields $\vec{n}_i (\tilde{\tau},\sigma) $,
$i=1,2$, parameterized as
\begin{equation}
\vec{n}_i = ( \cos \theta_i \cos \varphi_i , \cos \theta_i \sin
\varphi_i , \sin \theta_i )
\end{equation}

We now plug in $\dot{x}^-$, ${x^-}'$, $A$ and $B$ from
\eqref{xminussol} and \eqref{theA1}-\eqref{theB2} into the gauge
fixed Lagrangian
\begin{equation}
\CL_g = \CL - 2\pi  \kappa p_-  \dot{x}^-
\end{equation}
This gives
\begin{equation}
\CL_g = \CL_0 + \lambda' \CL_1 + {\lambda'}^2 \CL_2 + \cdots
\end{equation}
with
\begin{equation}
\label{cl0} \CL_0 = \frac{R^2}{16\pi^2} \sum_{i=1}^2 \Big[ \sin
\theta_i \dot{\varphi}_i - \pi^2 (\vec{n}_i')^2 \Big]
\end{equation}
\begin{align}
\CL_1 = \frac{R^2}{64 \pi^2} \Big[ \sum_{i=1}^2 \Big( 2
(\dot{\vec{n}}_i)^2 + \pi^4 (\vec{n}_i')^4 \Big) + 2 \pi^4
(\vec{n}_1')^2 (\vec{n}_2')^2  + 2 \psi_1 [ (\vec{n}_1')^2 -
(\vec{n}_2')^2 ] - \frac{\psi_1^2}{\pi^4} \Big]
\end{align}
\begin{align}
\CL_2 = & \frac{R^2}{64} \left\{ \frac{\psi_2}{\pi^4} \Big[
(\vec{n}_1)^2 -(\vec{n}_2)^2 - \frac{\psi_1}{\pi^4} \Big] - \frac{2
\psi_1'}{\pi^4 } - \frac{\psi_1}{2\pi^4}  \Big[ (\dot{\vec{n}}_1)^2
- (\dot{\vec{n}}_2)^2  + \pi^4 [ (\vec{n}_1')^4 - (\vec{n}_2')^4 ]
\Big] \right. \nn \\ & + \frac{\psi_1^2}{2\pi^4} [ (\vec{n}_1')^2 +
(\vec{n}_2')^2 ] - \frac{\pi^4}{2} [ (\vec{n}_1')^2 + (\vec{n}_2')^2
]^3  - 2 ( \dot{\vec{n}}_1 \cdot \vec{n}_1' + \dot{\vec{n}}_2 \cdot
\vec{n}_2')^2   \nn \\ & \left.  + [ (\dot{\vec{n}}_1)^2 +
(\dot{\vec{n}}_2)^2 ] [ ({\vec{n}_1}')^2 + ({\vec{n}_2}')^2 ]
\right\}
\end{align}
We see that $\CL_0$ is the sum of two Landau-Lifshitz models,
reproducing the result already found in \cite{Grignani:2008is}. In
$\CL_1$ we see that the first part is non-interacting in the two
$SU(2)$'s, then there is a interaction term and then a coupling to
$\psi$. We see that $\psi_1$ appears as a Lagrange-multiplier,
$i.e.$ it is not a dynamical field. The EOM for $\psi_1$ is found to
be satisfied provided
\begin{equation}
\label{psi1} \psi_1 = \pi^4 \Big[ (\vec{n}_1')^2  - (\vec{n}_2')^2
\Big]
\end{equation}
Inserting this into $\CL_1$, we get
\begin{equation}
\CL_1 = \frac{R^2}{32 \pi^2} \sum_{i=1}^2 \Big[ (\dot{\vec{n}}_i)^2
+ \pi^4 ({\vec{n}_i}')^4 \Big]
\end{equation}
We see that there are no interaction terms and the two $SU(2)$'s
appear symmetrically.

For $\CL_2$ we should first substitute in $\psi_1$ from
\eqref{psi1}. This gives
\begin{align}
\label{cl2} \CL_2 = & \frac{R^2}{64} \Big[ - 2 ( \dot{\vec{n}}_1
\cdot \vec{n}_1' + \dot{\vec{n}}_2 \cdot \vec{n}_2')^2 +
2(\dot{\vec{n}}_1)^2 ({\vec{n}_2}')^2 + 2(\dot{\vec{n}}_2)^2
({\vec{n}_1}')^2 \nn \\ & - \pi^4 \Big(
({\vec{n}_1}')^6+({\vec{n}_2}')^6 + ({\vec{n}_1}')^2
({\vec{n}_2}')^4+ ({\vec{n}_1}')^4 ({\vec{n}_2}')^2 + 8 ( \vec{n}_1'
\cdot \vec{n}_1'' - \vec{n}_2' \cdot \vec{n}_2'' )^2 \Big) \Big]
\end{align}
We see now that $\psi_2$ has disappeared from the Lagrangian after
substituting $\psi_1$. We also notice that there are interaction
terms in \eqref{cl2}.

We now want to eliminate the time derivatives in $\CL_1$ and
$\CL_2$. To do this we should perform a field redefinition,
following \cite{Kruczenski:2004kw}
\begin{equation}
\label{fieldred} \vec{n}_i \rightarrow \vec{n}_i + \lambda'
\vec{p}_i + {\lambda'}^2 \vec{q}_i
\end{equation}
in terms of $\vec{n}_i$ and their derivatives. By choosing
$\vec{p}_i$ and $\vec{q}_i$ it is possible to eliminate the
time-derivatives. Write first $\CL_1$ and $\CL_2$ as
\begin{equation}
\CL_1 = \sum_{i=1}^2 \vec{u}_i \cdot \frac{\delta \CL_0}{\delta
\vec{n}_i} + (\CL_1)_0 \spa \CL_2 = \sum_{i=1}^2 \vec{v}_i \cdot
\frac{\delta \CL_0}{\delta \vec{n}_i} + (\CL_2)_0
\end{equation}
where $(\CL_i)_0$ are $\CL_i$ without time-derivatives obtained by
using the leading EOM
\begin{equation}
\frac{\delta \CL_0}{\delta \vec{n}_i}=0
\end{equation}
One can check that we can use the same field redefinition as in
\cite{Kruczenski:2004kw}. This redefinition consists in choosing
$\vec{p}_i = - \vec{u}_i$ and $\vec{q}_i$ is furthermore chosen such
that we get the new Lagrangian
\begin{equation}
\label{newcl} \CL_g = \CL_0 + \lambda' (\CL_1)_0 + {\lambda'}^2
\widehat{\CL}_2
\end{equation}
with
\begin{equation}
\widehat{\CL}_2 = (\CL_2)_0 - \sum_{i=1}^2 \frac{\delta
(\CL_1)_0}{\delta \vec{n}_i} \cdot (\vec{u}_i)_0   + \sum_{i=1}^2
\sum_{a,b=1}^3 \left( \frac{\delta^2 \CL_0}{\delta (n_i)_a \delta
(n_i)_b}\right)_0 (u_{i,a})_0 (u_{i,b})_0
\end{equation}
Notice that the last two terms only involve $\CL_0$ and $\CL_1$.
Since $\vec{n}_1$ and $\vec{n}_2$ are decoupled in $\CL_0$ and
$\CL_1$ the last two terms do not contain any interaction terms
between the two two-spheres. This is also the reason why we can
directly use the field redefinition of \cite{Kruczenski:2004kw}.

The leading EOM is obtained from
\begin{equation}
\frac{\delta \CL_0}{\delta \vec{n}_i} = \frac{R^2}{16\pi^2 } \Big(
\vec{n}_i \times \dot{\vec{n}}_i + 2\pi^2 (\vec{n}''_i)_{\perp}
\Big)
\end{equation}
with
\begin{equation}
(\vec{n}''_i)_{\perp} = \vec{n}''_i + \vec{n} ( \vec{n}' )^2 \spa (
(\vec{n}''_i)_{\perp} )^2 = ( \vec{n}''_i )^2 - (\vec{n}'_i)^4
\end{equation}
Thus the leading EOM is
\begin{equation}
\vec{n}_i \times \dot{\vec{n}}_i = - 2\pi^2 (\vec{n}''_i)_{\perp}
\end{equation}
giving
\begin{equation}
(\dot{\vec{n}}_i)^2 = 4 \pi^4 ( (\vec{n}''_i)_{\perp} )^2 = 4 \pi^4
[ ( \vec{n}''_i )^2 - (\vec{n}'_i)^4 ] \spa \vec{n}_i' \cdot
\dot{\vec{n}}_i = - 2\pi^2 \vec{n}_i \cdot ( \vec{n}_i' \times
\vec{n}_i'' )
\end{equation}
This gives the following on-shell evaluations of $\CL_1$ and $\CL_2$
\begin{equation}
\label{newcl1} ( \CL_1 )_0 = \frac{\pi^2 R^2}{8}  \sum_{i=1}^2 \Big[
( \vec{n}'' _i)^2 - \frac{3}{4} (\vec{n}'_i )^4 \Big]
\end{equation}
\begin{align}
( \CL_2 )_0 = & \frac{\pi^4 R^2}{64} \Big\{ \sum_{i=1}^2 \Big( 7
(\vec{n}_i')^6 - 8 (\vec{n}'_i )^2(\vec{n}''_i )^2  \Big) + 8 [
(\vec{n}'_1 )^2(\vec{n}''_2 )^2 + (\vec{n}'_2 )^2(\vec{n}''_1 )^2 ]
\nn \\ & + 16 (\vec{n}_1' \cdot \vec{n}_1'')(\vec{n}_2' \cdot
\vec{n}_2'')  - 9 [ (\vec{n}_1')^2 (\vec{n}_2')^4 + (\vec{n}_2')^2
(\vec{n}_1')^4  ] \nn \\ & - 16 ( \vec{n}_1 \cdot ( \vec{n}'_1
\times \vec{n}''_1 ) ) ( \vec{n}_2 \cdot ( \vec{n}'_2 \times
\vec{n}''_2 ) ) \Big\}
\end{align}
We now compute
\begin{equation}
\widehat{\CL}_2 - ( \CL_2 )_0 = \frac{\pi^4 R^2}{2} \sum_{i=1}^2
\Big[ - (\vec{n}_i''')^2 + 2 (\vec{n}_i')^2 (\vec{n}_i'')^2 + 12 (
\vec{n}_i' \cdot \vec{n}_i'' )^2 - (\vec{n}_i')^6 \Big]
\end{equation}
Using this, we obtain
\begin{align}
\label{newcl2} \widehat{\CL}_2 = & \frac{\pi^4 R^2}{2} \Big\{
\sum_{i=1}^2 \Big( - (\vec{n}_i''')^2 - \frac{7}{4} (\vec{n}'_i
)^2(\vec{n}''_i )^2 + 12 ( \vec{n}_i' \cdot \vec{n}_i'' )^2 -
\frac{25}{32} (\vec{n}_i')^6 \Big) \nn \\ & + \frac{1}{4} [
(\vec{n}'_1 )^2(\vec{n}''_2 )^2 + (\vec{n}'_2 )^2(\vec{n}''_1 )^2 ]
- \frac{9}{32} [ (\vec{n}_1')^2 (\vec{n}_2')^4 + (\vec{n}_2')^2
(\vec{n}_1')^4  ] \nn \\ & + \frac{1}{2} (\vec{n}_1' \cdot
\vec{n}_1'')(\vec{n}_2' \cdot \vec{n}_2'')  - \frac{1}{2} (
\vec{n}_1 \cdot ( \vec{n}'_1 \times \vec{n}''_1 ) ) ( \vec{n}_2
\cdot ( \vec{n}'_2 \times \vec{n}''_2 ) ) \Big\}
\end{align}
The final sigma-model action is
\begin{equation}
\label{newact} I = \frac{4\pi J}{R^2} \int d\tilde{\tau} d\sigma
\Big[ \CL_0 + \lambda' (\CL_1)_0 + {\lambda'}^2 \widehat{\CL}_2
\Big]
\end{equation}
Thus, the action with time-derivatives only in the leading part is
given by \eqref{newact} along with \eqref{cl0}, \eqref{newcl1} and
\eqref{newcl2}. We notice again that for the leading part $\CL_0$,
corresponding to order $\lambda'$, and the first correction
$(\CL_1)_0$, corresponding to order ${\lambda'}^2$, there are no
interactions between the two two-spheres. In fact $\CL_0$ and
$(\CL_1)_0$ are equivalent to Lagrangians found for the $SU(2)$
sector of $\ads_5\times S^5$ in
\cite{Kruczenski:2003gt,Kruczenski:2004kw}. Instead for the second
order correction $\widehat{\CL}_2$, corresponding to $\lambda'^3$,
there are interactions between the two two-spheres, and also the
part acting only on a single $SU(2)$ is different from that found in
\cite{Kruczenski:2003gt,Kruczenski:2004kw} for the $SU(2)$ sector of
$\ads_5\times S^5$.

%We note here that the action \eqref{newc12} is supplemented with the
%level-matching condition
%
%\begin{equation}
%\label{momm} \sum_{i=1}^2 \int_0^{2\pi} d\sigma \sin \theta_i
%\varphi_i =0
%\end{equation}
%

\subsection{Computation of finite-size correction to energies}

In the following we compute the finite size correction to the
energies of the two string states $|s\rangle$ and $|t\rangle$
considered in Section \ref{sec:penrose} using the action
\eqref{newact}. In order to accomplish this, we first need to write
down the Hamiltonian. We begin by observing that the conjugate
momenta to $\varphi_i$ are
\begin{equation}
p_{\varphi_i} = J \sin \theta_i
\end{equation}
Notice that we have left out a factor $4\pi$ in front of the action.
With this, we can write the action \eqref{newact} as
\begin{equation}
I = \frac{J}{4\pi} \int d\tilde{\tau} d\sigma \Big[ \frac{1}{J}
\sum_{i=1}^2
 p_{\varphi_i} \dot{\varphi}_i - \CH_0 - \lambda' \CH_1 -
{\lambda'}^2 \CH_2 \Big]
\end{equation}
with
\begin{equation}
\CH_0 = \pi^2 \sum_{i=1}^2 (\vec{n}_i')^2 \spa \CH_1 = - 2\pi^4
\sum_{i=1}^2 \Big[ ( \vec{n}'' _i)^2 - \frac{3}{4} (\vec{n}'_i )^4
\Big]
\end{equation}
\begin{align}
\CH_2 = & 8 \pi^6 \left\{ \sum_{i=1}^2 \Big( (\vec{n}_i''')^2 +
\frac{7}{4} (\vec{n}'_i )^2(\vec{n}''_i )^2 - 12 ( \vec{n}_i' \cdot
\vec{n}_i'' )^2 + \frac{25}{32} (\vec{n}_i')^6 \Big) - \frac{1}{4} [
(\vec{n}'_1 )^2(\vec{n}''_2 )^2 + (\vec{n}'_2 )^2(\vec{n}''_1 )^2 ]
\right. \nn \\ & \left. + \frac{9}{32} [ (\vec{n}_1')^2
(\vec{n}_2')^4 + (\vec{n}_2')^2 (\vec{n}_1')^4  ] - \frac{1}{2}
(\vec{n}_1' \cdot \vec{n}_1'')(\vec{n}_2' \cdot \vec{n}_2'')  +
\frac{1}{2} ( \vec{n}_1 \cdot ( \vec{n}'_1 \times \vec{n}''_1 ) ) (
\vec{n}_2 \cdot ( \vec{n}'_2 \times \vec{n}''_2 ) ) \right\}
\end{align}
The Hamiltonian is thus
\begin{equation}
H = \frac{J}{4\pi} \int d\sigma \Big[ \CH_0 + \lambda' \CH_1 +
{\lambda'}^2 \CH_2 \Big]
\end{equation}
where $\vec{n}_i$ in terms of $\varphi_i$ and $p_{\varphi_i}$ is
\begin{equation}
\vec{n}_i = \Big( \sqrt{1 - \frac{p_{\varphi_i}^2}{J^2} } \cos
\varphi_i , \sqrt{1 - \frac{p_{\varphi_i}^2}{J^2} } \sin \varphi_i ,
\frac{p_{\varphi_i}}{J}  \Big)
\end{equation}

To compute the finite-size correction to a string state we want to
zoom in to $(\theta_i,\varphi_i)=(0,0)$. We do this by defining
\begin{equation}
x_i = \sqrt{J} \varphi_i \spa y_i = \sqrt{J} \theta_i
\end{equation}
The conjugate momenta for $x_i$ are
\begin{equation}
p_i = \sqrt{J} \sin \theta_i
\end{equation}
We now write the Hamiltonian up to $1/J^2$ corrections in terms of
the new variables. We get that
\begin{equation}
H = H_0 + \lambda' H_1 + {\lambda'}^2 H_2
\end{equation}
with
\begin{equation}
H_0=\frac{\pi}{4}\sum_{i=1}^2 \int_{0}^{2\pi} \left\{
{x_i'}^2+{p_i'}^2+ \frac{1}{J} \Big[
p_i^2\big({p_{i}'}^2-{x_i{'}}^2\big)\Big]\right\} \label{h0j}
\end{equation}
\begin{eqnarray}
H_1=-\frac{\pi^3}{2}\sum_{i=1}^2 \int_{0}^{2\pi} \left\{
{x_i''}^2+{p_{i}''}^2+ \frac{1}{J} \left[ (p_{i})^2 \big(
{p_{i}''}^2-{x_i''}^2\big) + 2 p_i p_i'' \left( {p_i'}^2 +
{x_i'}^2\right)-4 p_i p_i' x_i' x_i'' \right]\right\} \label{h1j}
\end{eqnarray}
\begin{eqnarray}
&&H_2=\pi^5\sum_{i=1}^2 \int_{0}^{2\pi}
\left\{2(x_i''')^2+2(p_{i}''')^2-
\frac{1}{2J}\left[4(p_i^2\left((x_i''')^2+{p_i'''}^2\right)+8(x_i')^3x_i'''
\right.\right.\nn \\[2mm]
&&+\left.\left.24p_ip_i'\left(x_i''x_i'''-p_i''p_i'''\right)+
24p_ix_i'\left(p_i''x_i'''-x_i''p_i'''\right)+24x_i'p_i'\left(p_i'x_i'''+x_i''p_i''
\right) \right.\right.\nn \\[2mm]
&&-7\left.\left((p_i')^2(x_i'')^2+(x_i')^2(p_i'')^2\right)
+5\left((x_i')^2(x_i'')^2+(p_i')^2(p_i'')^2\right)\right] \Big\}
+\bar{H}_{2} \label{h2j}
\end{eqnarray}
where $\bar{H}_{2}$ is given by
\begin{eqnarray}
\bar{H}_{2}=&&-\frac{\pi^5}{J}\int_0^{2\pi}\left\{ \frac{1}{2}\left[
(x_1'')^2+(p_1'')^2\right]\left[ (x_2')^2+(p_2')^2\right]+\left[
(x_2'')^2+(p_2'')^2\right]\left[
(x_1')^2+(p_1')^2\right]\right.\nn \\[2mm] && -\left(x_1''x_2''-p_1''p_2''\right)
\left(p_1'p_2'-x_1'x_2'\right)+\left(x_1''p_2''+p_1''x_2''\right)
\left(p_1'x_2'+x_1'p_2'\right)\Big\} \label{hint}
\end{eqnarray}
It is interesting to notice that the only part of the above
Hamiltonian with interactions between the two two-spheres is in
$\bar{H}_2$. This means that the leading interaction between the two
two-spheres appear at order ${\lambda'}^3/J$ in agreement with what
we have seen in Section \ref{sec:penrose}.

From the EOM we obtain the following mode expansions
\begin{equation}
x_i(t,\sigma)=\sum_{n=-\infty}^{\infty}\left(a_n^ie^{-i\bar\omega_nt+in\sigma}+a_n^{i\dagger}e^{i\bar\omega_n
t-in\sigma}\right) \label{coord2}
\end{equation}
\begin{equation}
p_i(t,\sigma)=-i\sum_{n=-\infty}^{\infty}\left(a_n^ie^{-i\bar\omega_nt+in\sigma}-a_n^{i\dagger}e^{i\bar\omega_n
t-in\sigma}\right) \label{coord3}
\end{equation}
where
\begin{equation}
\bar\omega_n= 2\pi^2(n^2-2\pi^2\lambda' n^4+8\pi^4\lambda'^2 n^6)
\label{freq}
\end{equation}
which coincides with the expansion up to $\mathcal{O}(\lambda'^4)$
of $\sqrt{\frac{1}{4}+2\pi^2n^2\lambda'}-\frac{1}{2}$. By imposing $
[a_m^i,(a_n^j)^\dagger] = \delta_{mn} \delta_{ij}$ we obtain the
standard canonical commutation relation
$[x_i(t,\sigma),p_{j}(t,\sigma')] = i\delta_{ij} \delta
(\sigma-\sigma')$.

From Eqs.~\eqref{h0j}-\eqref{h2j} we see that we obtain the free
spectrum
\begin{equation}
\label{E0} E_0 = 2\pi^2 \lambda' \sum_{n \in \Z} (n^2-2\pi^2\lambda'
n^4+8\pi^4\lambda'^2 n^6) (M^1_n + M^2_n) \spa \sum_{n\in \Z } n
(M^1_n + M^2_n) = 0
\end{equation}
which coincides with the expansion up to $\mathcal{O}(\lambda'^4)$
of the spectrum \eqref{penH}, \eqref{levelm}.

We now want to compute the $1/J$ corrections to the free spectrum.
These are obtained from the terms in Eqs.~\eqref{h0j}-\eqref{h2j}
which are quartic in the fields. Considering the state
$|s\rangle=a^{1\dagger}_na^{1\dagger}_{-n}|0\rangle$, we obtain
\begin{equation}
E-E_0=\frac{8 n^2 \pi^2 \lambda'}{J}-\frac{64 n^4 \pi^4
\lambda'^2}{J}+\frac{448 n^6 \pi^6\lambda^3}{J} \label{LLJ}
\end{equation}
which is in perfect agreement with the expansion of the energy
\eqref{Es} of the state $|s\rangle$ computed in Section
\ref{sec:penrose}. Moreover, considering the state
$|t\rangle=a^{1\dagger}_na^{2\dagger}_{-n}|0\rangle$, we obtain the
energy
\begin{equation}
E-E_0=-\frac{64 n^6 \pi^6\lambda'^3}{J} \label{LLJ2}
\end{equation}
This is also in perfect agreement with the expansion of the energy
\eqref{Et} of the state $|t\rangle$ computed in Section
\ref{sec:penrose}.

It is interesting to notice that the absence of interactions between
the two two-spheres at order $\lambda'^2$ here is due to the
non-trivial coupling with the non-dynamical field $\psi$, while in
Section \ref{sec:penrose} it is also due to the field $\psi$ but
there $\psi$ contributes through a second order perturbative
correction which is regularized using $\zeta$-function
regularization.

%%%%%%%%%%%%%%%%%%%%%%%%%%%%%%%%%%%%%%%%%%%%%%%%%%%%%%%%%%%%%
\section{Comparison with all-loop Bethe ansatz}
\label{sec:compare}

In the recent paper~\cite{Gromov:2008qe} Gromov and Vieira proposed
a set of all loop Bethe equations for the full asymptotic spectrum
of the $AdS_4/CFT_3$ duality. We shall provide here the explicit
expressions for the rapidities and the dressing factors for these
Bethe equations in the $SU(2)\times SU(2)$ sector in the strong
coupling regime, $\lambda \gg 1$. We shall then solve perturbatively
the Bethe equations constructed in this way and derive the first
non-trivial finite size corrections. These can then be compared to
the results we found from the explicit quantum calculations on two
oscillator states both from the string theory sigma model and from
the corresponding Landau-Lifshitz model.

For the $SU(2)\times SU(2)$ sector in the strong-coupling region
$\lambda \gg 1$ the Bethe equations read~\cite{Gromov:2008qe}
\begin{align}
e^{i p_k J}  = \prod_{j=1,j \ne k}^{K_p} S( p_k, p_j )
\prod_{j=1}^{K_p} \sigma( p_k,p_j) \prod_{j = 1}^{K_q} \sigma(
p_k,q_j)
\end{align}
\begin{align}
e^{i q_k J}  = \prod_{j=1,j \ne k}^{K_p} S( q_k, q_j )
\prod_{j=1}^{K_p} \sigma( q_k,q_j) \prod_{j = 1}^{K_q} \sigma(
q_k,p_j)
\end{align}
\begin{equation}
S(p_k,p_j) = \frac{\Phi( p_k )- \Phi(p_j) + i}{\Phi( p_k )-
\Phi(p_j) - i}
\end{equation}
The explicit form of the rapidities  $\Phi(p)$ and of the dressing
factor $\sigma( q_k,q_j)$ for this sector can be constructed along
the lines of those found in the $AdS_5/CFT_4$
duality~\cite{Beisert:2004hm,Arutyunov:2004vx, Beisert:2006ez}. The rapidities are
\begin{equation}
\Phi(p_j)=\cot{\frac{p_j}{2}}\sqrt{\frac{1}{4}+h(\lambda)\sin^2{\frac{p_j}{2}}}
\end{equation}
where, here, at strong coupling,
$h(\lambda)=2\lambda$~\cite{Gaiotto:2008cg,Grignani:2008is}. The relevant part of the 
dressing factor in terms of the conserved charges $Q_r(p)$ reads
\begin{equation}
\sigma(p_j,p_l)=\exp\left\{2i\sum_{r=0}^{\infty}\left(\frac{h(\lambda)}{16}\right)^{r+2}[Q_{r+2}(p_j)Q_{r+3}(p_l)-
Q_{r+2}(p_l)Q_{r+3}(p_j)]\right\}
\end{equation}
where we can write
\begin{equation}
Q_r(p_j)=\frac{2\sin{(\frac{r-1}{2}p_j})}{r-1}\left(\frac{\sqrt{\frac{1}{4}+
h(\lambda)\sin^2{\frac{p_j}{2}}}-\frac{1}{2}}{\frac{h(\lambda)}{4}\sin{\frac{p_l}{2}}}\right)^{r-1}
\end{equation}
We then have the dispersion relation
\begin{eqnarray}\label{disprel}
&&E=\Delta - J = \frac{h(\lambda)}{8} \left(\sum_{j =
1}^{K_p}Q_2(p_j)+\sum_{j = 1}^{K_q} Q_2(p_j)\right)\cr&&=\sum_{j =
1}^{K_p}\left( \sqrt{ \frac{1}{4} + h(\lambda)\sin^2 \frac{p_j}{2} }
-\frac{1}{2}\right)+ \sum_{j = 1}^{K_q}\left(  \sqrt{ \frac{1}{4} +
h(\lambda)\sin^2 \frac{q_j}{2} }-\frac{1}{2}\right)
\end{eqnarray}

We will now discuss the two magnon case in the $AdS_4\times \C P^3$
theory and solve the Bethe equations. These can be solved
perturbatively in $\lambda'$ and $J$ following the procedure adopted
for example in~\cite{Astolfi:2006is}. When one magnon is in one
$SU(2)$ sector and the other magnon is in the other~\footnote{This
situation corresponds to the state $|t\rangle$ on the string theory
side, so we label the corresponding energy/scaling dimension as
$E_t$.}, the scattering matrix becomes trivial and the momentum is
just given by the dressing phase. At the first non trivial order in
$\lambda'$ we get
\begin{equation}\label{2mixmag}
 e^{i p_1 J}=e^{i \frac{\lambda^2}{32}\left[Q_2(p_1)Q_3(q_1)-Q_3(p_1)Q_2(q_1)\right]}
\end{equation}
with $q_1=-p_1$ from the momentum constraint. Since the scattering
matrix is just 1 in this case, quite interestingly, the momentum
starts to receive corrections only at the order $\lambda'^2$ and
this will provide a non vanishing contribution to the finite size
correction to the energy only at the order $\lambda'^3$. This is
analogous to what we found on the string theory side both from
computing curvature corrections to the Penrose limit in Section
\ref{sec:penrose} and by considering a low energy expansion of the
string theory sigma model in Section \ref{sec:smexp}.

For the momentum we can consider an ansatz of the form
\begin{equation}\label{momentum2}
  p_1=\frac{2\pi n}{J}+\frac{a \lambda'^2}{J^2}+\CO\Big(\lambda'^3,\frac{1}{J^2}\Big)
\end{equation}
where $a$ is a parameter that will be determined by requiring that
the Bethe equations are satisfied at this order. Plugging
\eqref{momentum2} into \eqref{2mixmag} and expanding for small
$\lambda$ and large $J$ it is easy to determine $a$ as $a=-16\pi^5
n^5$. Using this result for the momentum in the dispersion relation
\eqref{disprel} we get for the first non trivial finite size
correction
\begin{equation}\label{mixedenergy}
  E_t=4n^2\pi^2\lambda'-8 n^4\pi^4{\lambda'}^2+32 n^6\pi^6{\lambda'}^3
  -\frac{64n^6\pi^6{\lambda'}^3}{J}+\CO\left(\frac{\lambda'}{J^2}\right)
\end{equation}
where we have written only the leading terms in
$\lambda'=\frac{\lambda}{J^2}$ and the first finite size correction.
This is provided by the last term. The result precisely coincides
with the one found in Sections \ref{sec:penrose} and
\ref{sec:smexp}, see eq.s \eqref{Et} and \eqref{LLJ2}. We see here
that the reason why the finite size correction only starts at three
loops is basically due to the fact that the only non trivial factor
in the Bethe equations is the dressing phase.

We solved the Bethe equations up to the order $\lambda'^8$ and we found perfect agreement with the string theory result for the energy $E_t$. We can then conclude that the dispersion relation up to the first order in $1/J$ for two magnons,
one in each $SU(2)$ sector, is
\begin{equation}\label{Etba}
E_t=2\sqrt{ \frac{1}{4} + 2\pi^2 n^2\lambda'  }
-1-\frac{\lambda'}{J}\frac{4 \pi^2n^2}{\frac{1}{4}+2\pi^2
n^2\lambda'}\left(\frac{1}{2}+2\pi^2 n^2\lambda'-\sqrt{ \frac{1}{4}
+ 2\pi^2n^2\lambda'  }\right)
\end{equation}
in the limit of large $\lambda$ with $\lambda' = \lambda/ J^2$
fixed.

The solution of the Bethe equations for the two magnon case, when
these belong to the same $SU(2)$ sector~\footnote{This situation
corresponds to the state $|s\rangle$ on the string theory side, so
we label the corresponding energy/scaling dimension as $E_s$.}, can
be obtained in the same way. The Bethe equations in this case read
\begin{align}
e^{i p_1J}  = \frac{\Phi( p_1 )- \Phi(p_2) + i}{\Phi( p_1 )-
\Phi(p_2) - i}~e^{i
\frac{\lambda^2}{32}\left[Q_2(p_1)Q_3(p_2)-Q_3(p_1)Q_2(p_2)\right]}
\label{be2}
\end{align}
where, from the momentum constraint, $p_2=-p_1$. The correct ansatz
for the expansion of the momentum now is
\begin{equation}\label{momentum}
  p_1=\frac{2\pi n}{J-1}+\frac{a \lambda'}{J^2}+\frac{b \lambda'^2}{J^2}+\CO\left(\lambda'^3,\frac{1}{J^2}\right)
\end{equation}
which substituted into the Bethe equations \eqref{be2} provides the
following solutions for the parameters $a$ and $b$: $a=-8\pi^3 n^3$,
$b=32 n^5 \pi^5$. Plugging the solution for the momentum back into
the dispersion relation we get
\begin{equation}\label{nonmixedenergy}
  E_s=4n^2\pi^2\lambda'-8 n^4\pi^4{\lambda'}^2+32 n^6\pi^6{\lambda'}^3+
  \frac{8n^2\pi^2\lambda'}{J}-\frac{64 n^4\pi^4{\lambda'}^2}{J}
 +\frac{448 n^6\pi^6{\lambda'}^3}{J}+\CO\left(\frac{\lambda'}{J^2}\right)
\end{equation}
where the last three terms give the finite size corrections which
coincide with those computed for this state in
sec.~\ref{sec:penrose} and \ref{sec:smexp}, see eq.s \eqref{Es} and
\eqref{LLJ}. Again we computed the finite-size corrections from the Bethe equations up to the order $\lambda'^8$ and we found perfect agreement with the string theory result for the energy $E_s$. We conclude that the
dispersion relation up to the first order in $\frac{1}{J}$ for two
magnons both in the same $SU(2)$ sector is
\begin{equation}\label{Esba}
E_s=2\sqrt{ \frac{1}{4} + 2\pi^2 n^2\lambda'  }
-1+\frac{\lambda'}{J}\frac{4 \pi^2 n^2}{\frac{1}{4}+2 \pi^2
n^2\lambda'}\left(\sqrt{ \frac{1}{4} + 2\pi^2n^2\lambda'  }-2\pi^2
n^2\lambda'\right)
\end{equation}
in the limit of large $\lambda$ with $\lambda' = \lambda/ J^2$
fixed.

In this section we have thus given evidence that the all loop Bethe
equations proposed in~\cite{Gromov:2008qe}, with the rapidities, the
dressing phase and the charges constructed here for the $SU(2)\times
SU(2)$ sector, are consistent with the finite size corrections
computed directly from the string sigma model and the corresponding
LL model.

%%%%%%%%%%%%%%%%%%%%%%%%%%%%%%%%%%%%%%%%%
\section*{Acknowledgments}

VGMP thanks O. Ohlsson Sax for useful discussions. TH and MO
thank the Carlsberg foundation for support. VGMP thanks VR for
partial financial support.

%The following two lines is for bibtex only:
%\bibliographystyle{C:/BIB/utphys}
%\bibliography{C:/BIB/mybib,C:/BIB/bibrot,C:/BIB/biblioniels}
%\bibliographystyle{../INPUT/newutphys_notitle}
%\bibliography{../BIB/mybib,../BIB/bibrot,../BIB/biblioniels}

\begin{thebibliography}{10}

\bibitem{Aharony:2008ug}
O.~Aharony, O.~Bergman, D.~L. Jafferis, and J.~Maldacena,
\href{http://arxiv.org/abs/0806.1218}{{\tt arXiv:0806.1218 [hep-th]}}.
%%CITATION = 0806.1218;%%.





\bibitem{Schwarz:2004yj}
J.~H. Schwarz,
\href{http://dx.doi.org/10.1088/1126-6708/2004/11/078}{{\em
  JHEP} {\bf 11} (2004)  078},
\href{http://arxiv.org/abs/hep-th/0411077}{{\tt
arXiv:hep-th/0411077}}.
%%CITATION = HEP-TH/0411077;%%.
%
%\bibitem{Gaiotto:2007qi}
D.~Gaiotto and X.~Yin,
  \href{http://dx.doi.org/10.1088/1126-6708/2007/08/056}{{\em JHEP} {\bf 08}
  (2007)  056},
\href{http://arxiv.org/abs/0704.3740}{{\tt arXiv:0704.3740
[hep-th]}}.
%%CITATION = 0704.3740;%%.
%
%\bibitem{Gaiotto:2008sd}
D.~Gaiotto and E.~Witten, \href{http://arxiv.org/abs/0804.2907}{{\tt
arXiv:0804.2907 [hep-th]}}.
%%CITATION = 0804.2907;%%.
%
%\bibitem{Bagger:2006sk}
J.~Bagger and N.~Lambert,
  \href{http://dx.doi.org/10.1103/PhysRevD.75.045020}{{\em Phys. Rev.} {\bf
  D75} (2007)  045020},
\href{http://arxiv.org/abs/hep-th/0611108}{{\tt
arXiv:hep-th/0611108}}.
%%CITATION = HEP-TH/0611108;%%.
%
%\bibitem{Gustavsson:2007vu}
A.~Gustavsson, \href{http://arxiv.org/abs/0709.1260}{{\tt
arXiv:0709.1260 [hep-th]}}.
%%CITATION = 0709.1260;%%.
%
%\bibitem{Bagger:2007jr}
J.~Bagger and N.~Lambert,
  \href{http://dx.doi.org/10.1103/PhysRevD.77.065008}{{\em Phys. Rev.} {\bf
  D77} (2008)  065008},
\href{http://arxiv.org/abs/0711.0955}{{\tt arXiv:0711.0955
[hep-th]}}.
%%CITATION = 0711.0955;%%.
%
%\bibitem{Bagger:2007vi}
J.~Bagger and N.~Lambert,
  \href{http://dx.doi.org/10.1088/1126-6708/2008/02/105}{{\em JHEP} {\bf 02}
  (2008)  105},
\href{http://arxiv.org/abs/0712.3738}{{\tt arXiv:0712.3738
[hep-th]}}.
%%CITATION = 0712.3738;%%.
%
%\bibitem{Bandres:2008vf}
M.~A. Bandres, A.~E. Lipstein, and J.~H. Schwarz,
  \href{http://dx.doi.org/10.1088/1126-6708/2008/05/025}{{\em JHEP} {\bf 05}
  (2008)  025},
\href{http://arxiv.org/abs/0803.3242}{{\tt arXiv:0803.3242
[hep-th]}}.
%%CITATION = 0803.3242;%%.
%
%\bibitem{VanRaamsdonk:2008ft}
M.~Van~Raamsdonk,
\href{http://dx.doi.org/10.1088/1126-6708/2008/05/105}{{\em
  JHEP} {\bf 05} (2008)  105},
\href{http://arxiv.org/abs/0803.3803}{{\tt arXiv:0803.3803
[hep-th]}}.
%%CITATION = 0803.3803;%%.
%
%\bibitem{Lambert:2008et}
N.~Lambert and D.~Tong, \href{http://arxiv.org/abs/0804.1114}{{\tt
arXiv:0804.1114 [hep-th]}}.
%%CITATION = 0804.1114;%%.
%
%\bibitem{Benvenuti:2008bt}
S.~Benvenuti, D.~Rodriguez-Gomez, E.~Tonni, and H.~Verlinde,
\href{http://arxiv.org/abs/0805.1087}{{\tt arXiv:0805.1087
[hep-th]}}.
%%CITATION = 0805.1087;%%.
%
%\bibitem{Bandres:2008kj}
M.~A. Bandres, A.~E. Lipstein, and J.~H. Schwarz,
\href{http://arxiv.org/abs/0806.0054}{{\tt arXiv:0806.0054
[hep-th]}}.
%%CITATION = 0806.0054;%%.
%
%\bibitem{Gomis:2008be}
J.~Gomis, D.~Rodriguez-Gomez, M.~Van~Raamsdonk, and H.~Verlinde,
\href{http://arxiv.org/abs/0806.0738}{{\tt arXiv:0806.0738
[hep-th]}}.
%%CITATION = 0806.0738;%%.
%
%\bibitem{Bagger:2008se}
J.~Bagger and N.~Lambert, \href{http://arxiv.org/abs/0807.0163}{{\tt
arXiv:0807.0163 [hep-th]}}.
%%CITATION = 0807.0163;%%.




\bibitem{Benna:2008zy}
M.~Benna, I.~Klebanov, T.~Klose, and M.~Smedback,
\href{http://arxiv.org/abs/0806.1519}{{\tt arXiv:0806.1519 [hep-th]}}.
%%CITATION = 0806.1519;%%.
%
%\bibitem{Ezhuthachan:2008ch}
B.~Ezhuthachan, S.~Mukhi, and C.~Papageorgakis,
\href{http://arxiv.org/abs/0806.1639}{{\tt arXiv:0806.1639 [hep-th]}}.
%%CITATION = 0806.1639;%%.
%
%\bibitem{Bhattacharya:2008bj}
J.~Bhattacharya and S.~Minwalla,
\href{http://arxiv.org/abs/0806.3251}{{\tt arXiv:0806.3251 [hep-th]}}.
%%CITATION = 0806.3251;%%.
%
%\bibitem{Honma:2008jd}
Y.~Honma, S.~Iso, Y.~Sumitomo, and S.~Zhang,
\href{http://arxiv.org/abs/0806.3498}{{\tt arXiv:0806.3498 [hep-th]}}.
%%CITATION = 0806.3498;%%.
%
%\bibitem{Imamura:2008nn}
Y.~Imamura and K.~Kimura,
\href{http://arxiv.org/abs/0806.3727}{{\tt arXiv:0806.3727 [hep-th]}}.
%%CITATION = 0806.3727;%%.
%
%\bibitem{Armoni:2008kr}
A.~Armoni and A.~Naqvi,
\href{http://arxiv.org/abs/0806.4068}{{\tt arXiv:0806.4068 [hep-th]}}.
%%CITATION = 0806.4068;%%.
%
%\bibitem{Hanany:2008qc}
A.~Hanany, N.~Mekareeya, and A.~Zaffaroni,
\href{http://arxiv.org/abs/0806.4212}{{\tt arXiv:0806.4212 [hep-th]}}.
%%CITATION = 0806.4212;%%.
%
%\bibitem{Hosomichi:2008jb}
K.~Hosomichi, K.-M. Lee, S.~Lee, S.~Lee, and J.~Park,
\href{http://arxiv.org/abs/0806.4977}{{\tt arXiv:0806.4977 [hep-th]}}.
%%CITATION = 0806.4977;%%.
%
%\bibitem{Beisert:2008qy}
N.~Beisert,
\href{http://arxiv.org/abs/0807.0099}{{\tt arXiv:0807.0099 [hep-th]}}.
%%CITATION = 0807.0099;%%.
%
%\bibitem{Terashima:2008sy}
S.~Terashima,
\href{http://arxiv.org/abs/0807.0197}{{\tt arXiv:0807.0197 [hep-th]}}.
%%CITATION = 0807.0197;%%.
%
%\bibitem{Terashima:2008ba}
S.~Terashima and F.~Yagi,
\href{http://arxiv.org/abs/0807.0368}{{\tt arXiv:0807.0368 [hep-th]}}.
%%CITATION = 0807.0368;%%.
%
%\bibitem{Ahn:2008gd}
C.~Ahn and P.~Bozhilov,
\href{http://arxiv.org/abs/0807.0566}{{\tt arXiv:0807.0566 [hep-th]}}.
%%CITATION = 0807.0566;%%.
%
%\bibitem{Chen:2008qq}
B.~Chen and J.-B. Wu,
\href{http://arxiv.org/abs/0807.0802}{{\tt arXiv:0807.0802 [hep-th]}}.
%%CITATION = 0807.0802;%%.
%
%\bibitem{Bandres:2008ry}
M.~A. Bandres, A.~E. Lipstein, and J.~H. Schwarz,
\href{http://arxiv.org/abs/0807.0880}{{\tt arXiv:0807.0880 [hep-th]}}.
%%CITATION = 0807.0880;%%.
%
%\bibitem{Gomis:2008vc}
J.~Gomis, D.~Rodriguez-Gomez, M.~Van~Raamsdonk, and H.~Verlinde,
\href{http://arxiv.org/abs/0807.1074}{{\tt arXiv:0807.1074 [hep-th]}}.
%%CITATION = 0807.1074;%%.
%
%\bibitem{Schnabl:2008wj}
M.~Schnabl and Y.~Tachikawa,
\href{http://arxiv.org/abs/0807.1102}{{\tt arXiv:0807.1102 [hep-th]}}.
%%CITATION = 0807.1102;%%.
%
%\bibitem{Li:2008ya}
T.~Li, Y.~Liu, and D.~Xie,
\href{http://arxiv.org/abs/0807.1183}{{\tt arXiv:0807.1183 [hep-th]}}.
%%CITATION = 0807.1183;%%.







\bibitem{Minahan:2008hf}
J.~A. Minahan and K.~Zarembo,
\href{http://arxiv.org/abs/0806.3951}{{\tt arXiv:0806.3951 [hep-th]}}.
%%CITATION = 0806.3951;%%.

\bibitem{Gaiotto:2008cg}
D.~Gaiotto, S.~Giombi, and X.~Yin,
\href{http://arxiv.org/abs/0806.4589}{{\tt arXiv:0806.4589 [hep-th]}}.
%%CITATION = 0806.4589;%%.

\bibitem{Grignani:2008is}
G.~Grignani, T.~Harmark, and M.~Orselli,
\href{http://arxiv.org/abs/0806.4959}{{\tt arXiv:0806.4959 [hep-th]}}.
%%CITATION = 0806.4959;%%.

\bibitem{Nishioka:2008gz}
T.~Nishioka and T.~Takayanagi,
\href{http://arxiv.org/abs/0806.3391}{{\tt arXiv:0806.3391 [hep-th]}}.
%%CITATION = 0806.3391;%%.

\bibitem{Grignani:2008te}
G.~Grignani, T.~Harmark, M.~Orselli, and G.~W. Semenoff,
\href{http://arxiv.org/abs/0807.0205}{{\tt arXiv:0807.0205 [hep-th]}}.
%%CITATION = 0807.0205;%%.

\bibitem{Hofman:2006xt}
D.~M. Hofman and J.~M. Maldacena, {\em J. Phys.} {\bf A39} (2006)
  13095--13118,
\href{http://arxiv.org/abs/hep-th/0604135}{{\tt
arXiv:hep-th/0604135}}.
%%CITATION = HEP-TH/0604135;%%.


\bibitem{Arutyunov:2006gs}
G.~Arutyunov, S.~Frolov, and M.~Zamaklar,
  \href{http://dx.doi.org/10.1016/j.nuclphysb.2006.12.026}{{\em Nucl. Phys.}
  {\bf B778} (2007)  1--35},
\href{http://arxiv.org/abs/hep-th/0606126}{{\tt
arXiv:hep-th/0606126}}.
%%CITATION = HEP-TH/0606126;%%.

\bibitem{Astolfi:2007uz}
D.~Astolfi, V.~Forini, G.~Grignani, and G.~W. Semenoff,
  \href{http://dx.doi.org/10.1016/j.physletb.2007.06.002}{{\em Phys. Lett.}
  {\bf B651} (2007)  329--335},
\href{http://arxiv.org/abs/hep-th/0702043}{{\tt
arXiv:hep-th/0702043}}.
%%CITATION = HEP-TH/0702043;%%.

\bibitem{Gromov:2008qe}
N.~Gromov and P.~Vieira,
\href{http://arxiv.org/abs/0807.0777}{{\tt arXiv:0807.0777 [hep-th]}}.
%%CITATION = 0807.0777;%%.

\bibitem{Arutyunov:2008if}
G.~Arutyunov and S.~Frolov,
\href{http://arxiv.org/abs/0806.4940}{{\tt arXiv:0806.4940 [hep-th]}}.
%%CITATION = 0806.4940;%%.

\bibitem{Stefanski:2008ik}
j.~Stefanski, B.,
\href{http://arxiv.org/abs/0806.4948}{{\tt arXiv:0806.4948 [hep-th]}}.
%%CITATION = 0806.4948;%%.

\bibitem{Fre:2008qc}
P.~Fre and P.~A. Grassi,
\href{http://arxiv.org/abs/0807.0044}{{\tt arXiv:0807.0044 [hep-th]}}.
%%CITATION = 0807.0044;%%.

\bibitem{Gromov:2008bz}
N.~Gromov and P.~Vieira,
\href{http://arxiv.org/abs/0807.0437}{{\tt arXiv:0807.0437 [hep-th]}}.
%%CITATION = 0807.0437;%%.

\bibitem{Beisert:2003ys}
N.~Beisert, {\em Nucl. Phys.} {\bf B682} (2004)  487--520,
\href{http://arxiv.org/abs/hep-th/0310252}{{\tt hep-th/0310252}}.
%%CITATION = HEP-TH 0310252;%%.
%
%\bibitem{Beisert:2005fw}
N.~Beisert and M.~Staudacher, {\em Nucl. Phys.} {\bf B727} (2005)  1--62,
\href{http://arxiv.org/abs/hep-th/0504190}{{\tt hep-th/0504190}}.
%%CITATION = HEP-TH 0504190;%%.
%
%\bibitem{Hernandez:2006tk}
R.~Hernandez and E.~Lopez, {\em JHEP} {\bf 07} (2006)  004,
\href{http://arxiv.org/abs/hep-th/0603204}{{\tt arXiv:hep-th/0603204}}.
%%CITATION = HEP-TH/0603204;%%.
%
%\bibitem{Beisert:2006qh}
N.~Beisert, {\em J. Stat. Mech.} {\bf 0701} (2007)  P017,
\href{http://arxiv.org/abs/nlin/0610017}{{\tt arXiv:nlin/0610017}}.
%%CITATION = NLIN/0610017;%%.
%
%
%\bibitem{Dorey:2007xn}
N.~Dorey, D.~M. Hofman, and J.~M. Maldacena,
  \href{http://dx.doi.org/10.1103/PhysRevD.76.025011}{{\em Phys. Rev.} {\bf
  D76} (2007)  025011},
\href{http://arxiv.org/abs/hep-th/0703104}{{\tt arXiv:hep-th/0703104}}.
%%CITATION = HEP-TH/0703104;%%.
%
\bibitem{Arutyunov:2004vx}
G.~Arutyunov, S.~Frolov, and M.~Staudacher, {\em JHEP} {\bf 10}
(2004)  016, \href{http://arxiv.org/abs/hep-th/0406256}{{\tt
hep-th/0406256}}.
%%CITATION = HEP-TH 0406256;%%.
\bibitem{Beisert:2006ez}
N.~Beisert, B.~Eden, and M.~Staudacher, {\em J. Stat. Mech.} {\bf 0701} (2007)
  P021,
\href{http://arxiv.org/abs/hep-th/0610251}{{\tt hep-th/0610251}}.
%
%%CITATION = HEP-TH/0610251;%%.
\bibitem{Callan:2003xr}
J.~Callan, Curtis~G. {\em et al.}, {\em Nucl. Phys.} {\bf B673} (2003)  3--40,
\href{http://arxiv.org/abs/hep-th/0307032}{{\tt hep-th/0307032}}.
%%CITATION = HEP-TH 0307032;%%.

\bibitem{Callan:2004uv}
J.~Callan, Curtis~G., T.~McLoughlin, and I.~J. Swanson, {\em Nucl. Phys.} {\bf
  B694} (2004)  115--169,
\href{http://arxiv.org/abs/hep-th/0404007}{{\tt hep-th/0404007}}.
%%CITATION = HEP-TH 0404007;%%.

\bibitem{Berenstein:2002jq}
D.~Berenstein, J.~M. Maldacena, and H.~Nastase, {\em JHEP} {\bf 04} (2002)
  013,
\href{http://arxiv.org/abs/hep-th/0202021}{{\tt hep-th/0202021}}.
%%CITATION = HEP-TH 0202021;%%.

\bibitem{Kruczenski:2003gt}
M.~Kruczenski, {\em Phys. Rev. Lett.} {\bf 93} (2004)  161602,
\href{http://arxiv.org/abs/hep-th/0311203}{{\tt hep-th/0311203}}.
%%CITATION = HEP-TH 0311203;%%.

\bibitem{Kruczenski:2004kw}
M.~Kruczenski, A.~V. Ryzhov, and A.~A. Tseytlin, {\em Nucl. Phys.} {\bf B692}
  (2004)  3--49,
\href{http://arxiv.org/abs/hep-th/0403120}{{\tt hep-th/0403120}}.
%%CITATION = HEP-TH/0403120;%%.

\bibitem{Bertolini:2002nr}
M.~Bertolini, J.~de~Boer, T.~Harmark, E.~Imeroni, and N.~A. Obers,
{\em JHEP}
  {\bf 01} (2003)  016,
\href{http://arxiv.org/abs/hep-th/0209201}{{\tt hep-th/0209201}}.
%%CITATION = HEP-TH 0209201;%%.

\bibitem{Sugiyama:2002tf}
K.~Sugiyama and K.~Yoshida,
  \href{http://dx.doi.org/10.1016/S0550-3213(02)00820-9}{{\em Nucl. Phys.} {\bf
  B644} (2002)  128--150},
\href{http://arxiv.org/abs/hep-th/0208029}{{\tt arXiv:hep-th/0208029}}.
%%CITATION = HEP-TH/0208029;%%.

\bibitem{Hyun:2002wu}
S.-j. Hyun and H.-j. Shin, {\em JHEP} {\bf 10} (2002)  070,
\href{http://arxiv.org/abs/hep-th/0208074}{{\tt arXiv:hep-th/0208074}}.
%%CITATION = HEP-TH/0208074;%%.

\bibitem{Minahan:2005mx}
J.~A. Minahan, A.~Tirziu, and A.~A. Tseytlin, {\em Nucl. Phys.} {\bf B735}
  (2006)  127--171,
\href{http://arxiv.org/abs/hep-th/0509071}{{\tt hep-th/0509071}}.
%%CITATION = HEP-TH/0509071;%%.

\bibitem{Astolfi:2008yw}
D.~Astolfi, G.~Grignani, T.~Harmark, and M.~Orselli,
\href{http://arxiv.org/abs/0804.3301}{{\tt arXiv:0804.3301 [hep-th]}}.
%%CITATION = 0804.3301;%%.

\bibitem{Harmark:2008gm}
T.~Harmark, K.~R. Kristjansson, and M.~Orselli,
\href{http://arxiv.org/abs/0806.3370}{{\tt arXiv:0806.3370 [hep-th]}}.
%%CITATION = 0806.3370;%%.

\bibitem{Beisert:2004hm}
N.~Beisert, V.~Dippel, and M.~Staudacher, {\em JHEP} {\bf 07} (2004)  075,
\href{http://arxiv.org/abs/hep-th/0405001}{{\tt hep-th/0405001}}.
%%CITATION = HEP-TH 0405001;%%.

\bibitem{Astolfi:2006is}
D.~Astolfi, V.~Forini, G.~Grignani, and G.~W. Semenoff, {\em JHEP} {\bf 09}
  (2006)  056,
\href{http://arxiv.org/abs/hep-th/0606193}{{\tt hep-th/0606193}}.
%%CITATION = HEP-TH/0606193;%%.

\end{thebibliography}

\providecommand{\href}[2]{#2}\begingroup\raggedright\endgroup

\end{document}